\documentclass{emulateapj}

\newcommand{\msun}{{\rm M}_{\odot}}
\newcommand{\rsun}{{\rm R}_{\odot}}
\newcommand{\lsun}{{\rm L}_{\odot}}

\newcommand{\kms}{\rm km\ s^{-1}}
\newcommand{\kelvin}{\rm K}

\newcommand{\angstrom}{\rm \AA}

\begin{document}

\shorttitle{GJ~3236}
\shortauthors{Irwin et al.}

\title{GJ~3236: a new bright, very low-mass eclipsing binary system
  discovered by the MEarth observatory}

\author{Jonathan~Irwin, David~Charbonneau, Zachory~K.~Berta,
  Samuel~N.~Quinn, David~W.~Latham, Guillermo~Torres, Cullen~H.~Blake,
  Christopher~J.~Burke, Gilbert~A.~Esquerdo, G\'abor F\"ur\'esz,
  Douglas~J.~Mink, Philip~Nutzman and Andrew~H.~Szentgyorgyi}
\affil{Harvard-Smithsonian Center for Astrophysics, 60 Garden St.,
  Cambridge, MA 02138, USA}
\email{jirwin@cfa.harvard.edu}

\author{Michael~L.~Calkins and Emilio~E.~Falco}
\affil{Fred Lawrence Whipple Observatory, Smithsonian Astrophysical Observatory, 670 Mount Hopkins Road,
  Amado, AZ 85645, USA}

\author{Joshua~S.~Bloom\altaffilmark{1} and Dan~L.~Starr}
\affil{Astronomy Department, University of California, 445 Campbell
  Hall, Berkeley, CA 94720, USA}

\altaffiltext{1}{Sloan Research Fellow}

\begin{abstract}
We report the detection of eclipses in GJ~3236, a bright ($I =
11.6$) very low mass binary system with an orbital period of $0.77\
{\rm days}$.  Analysis of light- and radial velocity curves of 
the system yielded component masses of $0.38 \pm 0.02$ and $0.28 \pm
0.02\ \msun$.  The central values for the stellar radii are larger
than the theoretical models predict for these masses, in agreement
with the results for existing eclipsing binaries, although the present
$5\%$ observational uncertainties limit the significance of the
larger radii to approximately $1 \sigma$.  Degeneracies in the light
curve models resulting from the unknown configuration of surface spots on
the components of GJ~3236 currently dominate the uncertainties in the
radii, and could be reduced by obtaining precise, multi-band
photometry covering the full orbital period.  The system appears to be
tidally synchronized and shows signs of high activity levels as
expected for such a short orbital period, evidenced by strong
H$\alpha$ emission lines in the spectra of both components.  These
observations probe an important region of mass-radius parameter space
around the predicted transition to fully-convective stellar interiors,
where there are a limited number of precise measurements available in
the literature.
\end{abstract}


\section{Introduction}
\label{intro_sect}

Detached, double-lined eclipsing binaries provide a largely
model-independent means to precisely and accurately measure
fundamental stellar properties, particularly masses and radii.  In the
best-observed systems the precision of these can be at the $< 1$ per
cent level, and thus place stringent constraints on stellar evolution
models (e.g. \citealt{andersen1991}).

Despite this, as far as we are aware, there are at present
only four known systems with one or more components between $0.4\
\msun$ and the hydrogen burning limit on the main sequence (i.e. old,
field stars): CM~Dra
\citep{eggen1967,lacy1977,metcalfe1996,morales2009}, CU~Cnc~B
\citep{delfosse1999,ribas2003}, LP~133-373 \citep{vaccaro2007}, and
SDSS~J031824-010018 \citep{blake2008}.  Although JW~380
\citep{irwin2007a}, NSVS~02502726B \citep*{cakirli2009}, and the
NGC~1647 system of \citet{hebb2006} also satisfy this mass criterion,
these objects are still contracting on the pre--main-sequence.

Of the known systems, the only two with parameters determined to
better than $2$ per cent are CM~Dra and CU~Cnc.  For
SDSS~J031824-010018, the knowledge of the parameters is limited
largely by radial velocity errors, since this system is extremely
faint (SDSS $r = 19.3$) and has an extremely short orbital period
($0.41\ {\rm days}$), meaning long integrations cannot be used to
obtain better signal-to-noise.  Hence there is little possibility for
substantial improvement in the parameters in the near future.  It is
clear that, in order to better-constrain the stellar mass-radius
relation on the main sequence, more bright, low-mass eclipsing
binaries are needed to yield extremely precise masses and radii.

Furthermore, observations of these systems have indicated significant
discrepancies with the stellar models.  This is particularly the case
below $0.4\ \msun$, and the components of CM~Dra (the lowest-mass
system with $<1\%$ observational errors) have radii $10-15$ per
cent larger than the theoretical predictions from state-of-the-art
stellar evolution models.  It has been suggested
(e.g. \citealt{chabrier2007}) that the reason for this discrepancy may
be that close binaries are not actually representative of single-stars
at this level of precision.  The effect of the close companion and
tidal locking is likely to significantly increase activity levels in
close binaries, and these authors suggest that it could be this effect
that is responsible for the inflated radii of CM~Dra.

\citet{lopezmorales2007} examined the available sample of single-star
and eclipsing binary measurements to search for correlations of radius
with activity levels and metallicity, finding that such a correlation
of activity with radius does appear to exist for members of close
binary stars.  This result was based on a small number of
measurements for late M-dwarfs, and the conclusions would be
strengthened by the availability of additional precise estimates of
the parameters for binaries with a range of activity levels and
orbital periods, to explore the available parameter space.

We present the discovery of a new bright, low-mass eclipsing binary
system, GJ 3236.  The components show high activity levels, as
evidenced by H$\alpha$ emission lines in the spectra of both stars,
and show rotational modulations in the light curves that are
synchronized with the binary orbital period of $0.77\ {\rm
  days}$.  This system thus has the potential to yield an additional
precise test of stellar evolution models for high activity and short
orbital periods.

\section{Observations and data reduction}
\label{obs_sect}

\subsection{MEarth photometry}
\label{mearth_obs_sect}

Eclipses in GJ 3236 were detected during 2008 January in the first two
weeks of routine operations of the newly-commissioned MEarth
observatory, a system designed primarily to search for transiting
super-Earth exoplanets orbiting around the nearest $2000$ mid to
late M-dwarfs in the northern hemisphere (\citealt{nutzman2008};
\citealt{irwin2008}).  Exposure times on each field observed by MEarth
are tailored to achieve sensitivity to a particular planet size for
the assumed stellar parameters of the target star, and were $82\ {\rm
  s}$ for GJ 3236.

After the initial detection of the eclipses, we switched to a
follow-up mode, observing at the highest possible cadence
(i.e. continuously, resulting in a cadence of approximately $2\ {\rm
minutes}$ including overheads) during eclipse and for an additional
$1\ {\rm hour}$ window surrounding the eclipse, and at the normal
cadence for the MEarth survey of approximately $20\ {\rm minutes}$ for
the remainder of the time.  In this way, we sample as well as possible
the rapid flux decrement during eclipse, and the out of eclipse
portions of the light curve.  We note that sampling the latter is of
crucial importance for obtaining accurate parameters from light curve
analyses of binary systems showing out of eclipse modulations, such as
the present one.

We obtained data both in the initial season after the eclipses were
first detected (2008 January to May, inclusive), and following the
summer monsoon (2008 October onwards).  During this time, there were a
number of software improvements (including the ability to update the
telescope pointing after taking each science image, resulting in
substantially improved light curves due to minimizing the drift across
the detector, and hence the effect of flat fielding errors, fringing,
etc.) and our filter system was changed from Cousins $I$ to a custom
long-pass filter with transmission from $715\ {\rm nm}$, and limited
at the long wavelength end by the tail of the CCD quantum efficiency
curve (see \citealt{nutzman2008}).  We have elected to use only the
latter data-set for modeling, despite this having the disadvantage of
being observed in a non-standard bandpass, because it is approximately
contemporaneous with our $V$-band follow-up photometry.  We see
evidence for evolution of the amplitude and phase of the out of
eclipse modulation between the two observing seasons, so this
simplifies the analysis by allowing the use of a single set of spot
parameters to describe both bandpasses.

A total of $1540$ observations were obtained between UT 2008 October 2
and UT 2008 December 11, including 6 primary and 3 secondary eclipses
sampled at high cadence.

Data were reduced using the standard MEarth reduction pipeline, which
is at present identical to the Monitor project pipeline described in
\citet{irwin2007b}.  We used a smaller aperture radius of $5\ {\rm
  pixels}$ ($3\farcs8$) than the usual value of $10\ {\rm pixels}$
adopted for bright stars, due to the presence of a nearby star
approximately $7\arcsec$ from GJ~3236 at the epoch of our
observations, and $4\ {\rm mag}$ fainter in the MEarth bandpass.  This
star does not share a common proper motion with GJ~3236 and is
therefore not physically associated.  Reducing the aperture size in
this way reduces the flux contributed by this star in the light curve
of GJ~3236 to a negligible level (we estimate a $0.1\%$ error in the
measured fluxes, and thus also in the eclipse depths).

\begin{figure}
\centering
\includegraphics[angle=270,width=2.8in]{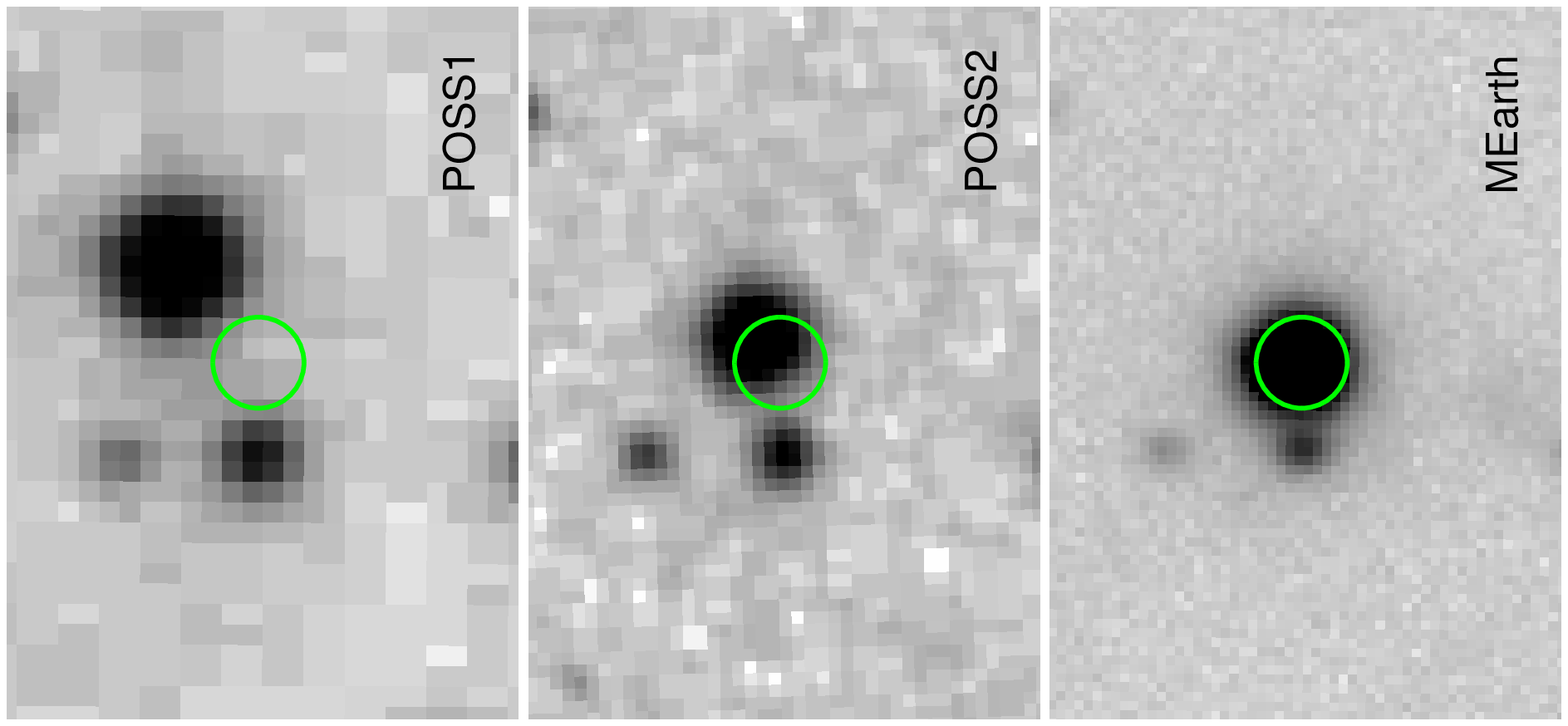}
\caption{Images of GJ~3236 centered on its position as measured from
  the MEarth data.  The circle shows the approximate position and
  size of the $5\arcsec$ (radius) photometric aperture used to derive
  our light curves.  Data are from the first and second epoch Palomar
  sky surveys as provided by the Digitized Sky Survey (top and center
  panels), and the MEarth stacked master image (bottom panel).  The
  approximate epochs of the images are $1954.1$ (POSS-1), $1992.7$
  (POSS-2), and $2008.8$ (MEarth).  All three panels have the same
  center, scale and alignment on-sky, with north up and east to the
  left, covering $1\arcmin$ in the horizontal direction.}
\label{third_light_images}
\end{figure}

The high proper motion of the GJ~3236 system allows us to constrain
the contribution of any additional background stars in the photometric
aperture that are not co-moving with GJ~3236 itself, using previous
epochs of imaging.  We show in Figure \ref{third_light_images} a
series of three images centered on the position of the photometric
aperture used for GJ~3236 in the MEarth images, demonstrating that
there are no such objects.  If any additional third light is
indeed present, it must therefore be co-moving with GJ~3236, meaning
it is highly likely to be physically associated, given the extremely
small probability of such a chance alignment.

The photometry was calibrated using observations of standard star
fields from \citet{landolt1992} taken automatically each night by the
MEarth observation scheduling software.  By fitting all the
standard star observations from several photometric nights, we derived
the following color equation to convert between the MEarth bandpass
and the standard Johnson-Cousins system:
\begin{equation}
{\rm MEarth} = I_{\rm C} - 0.18\ \left(V_{\rm J} - I_{\rm C}\right)
\label{color_eq}
\end{equation}

The full MEarth light curve is reproduced in Table
\ref{mearth_lc_table}.

\begin{deluxetable*}{lrrrrrrr}
\tabletypesize{\normalsize}
\tablecaption{\label{mearth_lc_table} MEarth light curve of GJ~3236.}
\tablecolumns{8}
\tablewidth{6in}

\tablehead{\colhead{HJD} & \colhead{MEarth} & \colhead{Error\tablenotemark{a}} &
    \colhead{$\Delta m$\tablenotemark{b}} & \colhead{FWHM
    (pix)\tablenotemark{c}} & \colhead{Airmass} &
    \colhead{$x$ (pix)\tablenotemark{d}} & \colhead{$y$ (pix)\tablenotemark{d}}
}

\startdata
$2454741.691783$ &$11.1329$ &$0.0036$ &$-0.278$ &$ 7.24$ &$1.94119$ &$1029.66$ &$1048.57$ \\
$2454741.704376$ &$11.1151$ &$0.0035$ &$-0.337$ &$ 7.18$ &$1.85590$ &$1030.44$ &$1049.71$ \\
$2454741.716182$ &$11.1159$ &$0.0035$ &$-0.223$ &$ 6.78$ &$1.78328$ &$1030.33$ &$1048.66$ \\
$2454741.727410$ &$11.1127$ &$0.0035$ &$-0.209$ &$ 6.81$ &$1.72032$ &$1030.19$ &$1048.31$ \\
$2454741.738267$ &$11.1180$ &$0.0035$ &$-0.226$ &$ 6.95$ &$1.66467$ &$1030.41$ &$1048.00$ \\
\enddata

\tablenotetext{a}{Estimated using a standard CCD noise model,
  including contributions from Poisson noise in the stellar counts, sky
  noise, readout noise and errors in the sky background estimation.}
\tablenotetext{b}{Correction to the frame magnitude zero-point applied
  in the differential photometry procedure.  More negative numbers
  indicate greater losses.}
\tablenotetext{c}{Median FWHM of the stellar images on the frame.  The
  plate scale was $0\farcs757/{\rm pix}$.}
\tablenotetext{d}{$x$ and $y$ pixel coordinates of GJ~3236 on the CCD
  image, derived using a standard intensity-weighted moments
  analysis.}

\tablecomments{Table \ref{mearth_lc_table} is published in its
  entirety in the electronic edition of the Astrophysical Journal.  A
  portion is shown here for guidance regarding its form and content.}

\end{deluxetable*}

\subsection{FLWO $1.2\ {\rm m}$ $V$-band follow-up photometry}
\label{v_sect}

Observations centered around the primary eclipse of UT 2008 September
25 and the secondary eclipse of UT 2008 September 27 were obtained
using the KeplerCam instrument on the FLWO $1.2\ {\rm m}$ telescope.
We used the standard binning $2 \times 2$ readout mode, since the
plate scale of $0\farcs34$ per unbinned pixel significantly
oversamples the typical seeing at FLWO.  The resulting plate scale was
$0\farcs67$ per summed pixel.  We used the $V$ filter and an
exposure time of $120\ {\rm s}$.  We also used observations of
standard star fields from \citet{landolt1992} taken on UT 2008
September 28 to calibrate the KeplerCam photometry onto the standard
Johnson-Cousins system.  A total of $245$ observations were taken,
where possible starting $2\ {\rm hours}$ before first contact and
finishing $2\ {\rm hours}$ after last contact to sample the
out-of-eclipse portions of the light curve and thus allow the eclipse
depths to be properly measured.  The end of the primary eclipse was
missed due to twilight.

These photometric data were reduced using the same pipeline as
described in \S \ref{mearth_obs_sect}.  We used an aperture radius of
$5$ binned pixels, corresponding to $3\farcs4$ on-sky (recalling from
\S \ref{mearth_obs_sect} that the size of this aperture is important
due to the nearby star).  The difference in magnitude between GJ~3236
and the fainter star is only $2.3\ {\rm mag}$ in $V$-band, which leads
to approximately a $0.6\%$ error in the measured fluxes of GJ~3236,
and thus a similar error in the measured eclipse depths.  Since the
point spread functions on our images are not very well-behaved it will
be challenging to reduce this using PSF-fitting photometry, and it is
still substantially smaller than the other sources of error in the
present light curve models, so we have not pursued this issue further.

The full $V$-band light curve is reproduced in Table \ref{v_lc_table}.

\begin{deluxetable*}{lrrrrrrr}
\tabletypesize{\normalsize}
\tablecaption{\label{v_lc_table} $V$-band light curve of GJ~3236.}
\tablecolumns{8}
\tablewidth{6in}

\tablehead{\colhead{HJD} & \colhead{$V$} & \colhead{Error\tablenotemark{a}} &
    \colhead{$\Delta m$\tablenotemark{b}} & \colhead{FWHM
    (pix)\tablenotemark{c}} & \colhead{Airmass} &
    \colhead{$x$ (pix)\tablenotemark{d}} & \colhead{$y$ (pix)\tablenotemark{d}}
}

\startdata
$2454734.882776$ &$14.2717$ &$0.0030$ &$-0.137$ &$ 6.29$ &$1.30113$ &$ 953.69$ &$1372.35$ \\
$2454734.884327$ &$14.2676$ &$0.0029$ &$-0.094$ &$ 6.08$ &$1.29916$ &$ 953.59$ &$1372.41$ \\
$2454734.885866$ &$14.2672$ &$0.0029$ &$-0.068$ &$ 5.92$ &$1.29726$ &$ 953.64$ &$1372.36$ \\
$2454734.887406$ &$14.2698$ &$0.0029$ &$-0.112$ &$ 6.14$ &$1.29541$ &$ 953.69$ &$1372.47$ \\
$2454734.888957$ &$14.2696$ &$0.0029$ &$-0.093$ &$ 6.05$ &$1.29360$ &$ 953.66$ &$1372.63$ \\
\enddata

\tablenotetext{a}{Estimated using a standard CCD noise model,
  including contributions from Poisson noise in the stellar counts, sky
  noise, readout noise and errors in the sky background estimation.}
\tablenotetext{b}{Correction to the frame magnitude zero-point applied
  in the differential photometry procedure.  More negative numbers
  indicate greater losses.}
\tablenotetext{c}{Median FWHM of the stellar images on the frame.  The
  plate scale was $0\farcs67/{\rm pix}$.}
\tablenotetext{d}{$x$ and $y$ pixel coordinates of GJ~3236 on the CCD
  image, derived using a standard intensity-weighted moments
  analysis.}

\tablecomments{Table \ref{v_lc_table} is published in its
  entirety in the electronic edition of the Astrophysical Journal.  A
  portion is shown here for guidance regarding its form and content.}

\end{deluxetable*}

\subsection{PAIRITEL $J$-band follow-up photometry}
\label{j_sect}

Observations in the near-infrared $J$-band were obtained using the
robotic Peters automated infrared imaging telescope (PAIRITEL) from UT
2008 February 17 to UT 2008 March 3 (inclusive).  We scheduled
observations around the times of primary and secondary eclipse,
obtaining $457$ data points spread over $10\ {\rm nights}$.  Exposure
times were $7.8\ {\rm s}$.  The observations were automatically
scheduled, collected, and reduced by the fully robotic PAIRITEL
observing system \citep{bloom2006,blake2005}.  We then produced
differential photometry using a set of comparison stars chosen from
the 2MASS catalog.  We estimate a photometric error of approximately
$2\%$ per data point from the scatter of the out-of-eclipse
measurements.  The full $J$-band light curve is reproduced in Table
\ref{j_lc_table}.

\begin{deluxetable}{lrrrrrrr}
\tabletypesize{\normalsize}
\tablecaption{\label{j_lc_table} $J$-band light curve of GJ~3236.}
\tablecolumns{3}
\tablewidth{3.2in}

\tablehead{\colhead{HJD} & \colhead{$J$} & \colhead{Error\tablenotemark{a}}
}

\startdata
$2454513.6413$ &$10.015$ &$0.020$ \\
$2454513.6421$ &$10.040$ &$0.020$ \\
$2454513.6430$ &$10.052$ &$0.020$ \\
$2454513.6438$ &$10.078$ &$0.020$ \\
$2454513.6447$ &$10.042$ &$0.020$ \\
\enddata

\tablenotetext{a}{Equal observational errors of $0.02$ mag were
  assumed on each data point.}

\tablecomments{Table \ref{j_lc_table} is published in its
  entirety in the electronic edition of the Astrophysical Journal.  A
  portion is shown here for guidance regarding its form and content.}

\end{deluxetable}

\subsection{Spectroscopy}
\label{spec_sect}

Spectroscopic observations were obtained using the TRES fiber-fed
\'echelle spectrograph on the FLWO $1.5\ {\rm m}$ Tillinghast
reflector.  We used the medium fiber ($2\farcs3$ projected diameter) 
throughout, yielding a resolving power of $R \simeq 48\,000$.

Table \ref{spec_obs_table} summarizes these measurements, including
the radial velocities of both components of the binary derived
therefrom.  The TRES instrument is extremely stable, so we acquired
ThAr wavelength calibration exposures before or after the target
exposures, rather than using the simultaneous calibration fiber, which
can lead to contamination of the target spectrum in the red-most
orders from the very strong Ar lines in the ThAr lamp spectrum.  A
second fiber was placed on sky, but this was not used due to a
problem causing extremely poor throughput in the TRES sky fiber, which
has since been resolved.

\begin{deluxetable*}{lrrrrrr}
\tabletypesize{\normalsize}
\tablecaption{\label{spec_obs_table} Barycentric radial velocity measurements of GJ~3236.}
\tablecolumns{7}

\tablehead{
              & \multicolumn{2}{c}{Absorption (Barnard's star)\tablenotemark{a}} & \multicolumn{2}{c}{Emission (GJ~856A)} & \multicolumn{2}{c}{Emission (GJ~3379)} \\
\colhead{HJD} & \colhead{$v_1$ ($\kms$)} & \colhead{$v_2$ ($\kms$)} & \colhead{$v_1$ ($\kms$)} & \colhead{$v_2$ ($\kms$)} & \colhead{$v_1$ ($\kms$)} & \colhead{$v_2$ ($\kms$)}
}

\startdata
$2454755.9072$ &           &           & $ -44.23$ & $  86.66$ & $ -42.91$ & $  87.65$ \\
$2454755.9154$ &           &           & $ -49.99$ & $  92.02$ & $ -48.28$ & $  91.13$ \\
$2454755.9241$ &           &           & $ -53.77$ & $ 100.02$ & $ -52.82$ & $  99.68$ \\
$2454757.8115$ & $  55.29$ & $ -37.60$ & $  55.70$ & $ -44.31$ & $  56.17$ & $ -42.63$ \\
$2454757.8192$ & $  58.96$ & $ -49.14$ & $  61.49$ & $ -51.70$ & $  62.63$ & $ -51.13$ \\
$2454757.8272$ & $  61.92$ & $ -56.41$ & $  65.03$ & $ -57.34$ & $  66.39$ & $ -57.10$ \\
$2454757.8352$ & $  66.54$ & $ -66.09$ & $  70.44$ & $ -63.70$ & $  71.87$ & $ -62.77$ \\
$2454756.8324$ &           &           & $ -66.67$ & $ 118.68$ & $ -66.39$ & $ 119.91$ \\
$2454756.8406$ &           &           & $ -64.81$ & $ 114.28$ & $ -63.84$ & $ 116.20$ \\
$2454756.8483$ &           &           & $ -63.87$ & $ 114.44$ & $ -62.96$ & $ 117.50$ \\
$2454729.8825$ &           &           & $ -50.95$ & $  96.13$ & $ -49.89$ & $  96.98$ \\
$2454729.8883$ & $ -53.26$ & $  81.21$ & $ -48.21$ & $  90.02$ & $ -47.42$ & $  91.53$ \\
$2454729.8941$ & $ -50.49$ & $  79.39$ & $ -44.90$ & $  87.55$ & $ -44.18$ & $  88.84$ \\
$2454729.8999$ & $ -46.09$ & $  84.13$ & $ -40.92$ & $  81.58$ & $ -40.06$ & $  83.26$ \\
$2454729.9057$ & $ -38.67$ & $  74.80$ & $ -40.44$ & $  77.21$ & $ -39.52$ & $  79.13$ \\
$2454759.9137$ &           &           & $ -59.96$ & $ 119.56$ & $ -58.57$ & $ 123.32$ \\
$2454759.9252$ &           &           & $ -61.53$ & $ 104.53$ & $ -55.20$ & $ 102.53$ \\
$2454759.9513$ &           &           & $ -57.89$ & $ 106.49$ & $ -56.91$ & $ 107.42$ \\
$2454758.8510$ & $  51.55$ & $ -31.67$ & $  48.34$ & $ -37.76$ & $  48.98$ & $ -35.47$ \\
$2454758.8587$ & $  48.84$ & $ -26.92$ & $  44.02$ & $ -32.35$ & $  44.54$ & $ -30.38$ \\
$2454758.8668$ & $  48.57$ & $ -21.32$ & $  39.19$ & $ -21.43$ & $  40.09$ & $ -18.55$ \\
$2454758.8745$ & $  38.38$ & $ -17.90$ & $  33.12$ & $ -16.09$ & $  34.85$ & $ -12.88$ \\
$2454730.9731$ & $  90.88$ & $-101.58$ & $ 100.49$ & $-101.59$ & $ 102.23$ & $-100.68$ \\
$2454730.9768$ & $  79.42$ & $-110.01$ & $ 101.24$ & $ -96.71$ & $ 102.05$ & $ -96.08$ \\
$2454730.9806$ & $  91.46$ & $-105.07$ & $  99.63$ & $ -93.68$ & $ 100.75$ & $ -93.28$ \\
$2454730.9843$ & $  90.40$ & $-110.53$ & $  99.04$ & $ -97.74$ & $ 100.43$ & $ -97.25$ \\
$2454730.9880$ & $  87.48$ & $ -96.16$ & $  97.66$ & $ -96.19$ & $  99.13$ & $ -95.82$ \\
$2454731.9797$ & $ -42.46$ & $  73.24$ & $ -34.37$ & $  74.11$ & $ -33.33$ & $  74.37$ \\
$2454731.9835$ & $ -39.77$ & $  73.55$ & $ -36.22$ & $  76.30$ & $ -35.22$ & $  77.22$ \\
$2454731.9872$ & $ -46.24$ & $  83.24$ & $ -37.64$ & $  78.70$ & $ -36.75$ & $  80.16$ \\
$2454731.9909$ & $ -30.10$ & $  76.54$ & $ -40.50$ & $  83.11$ & $ -39.25$ & $  84.72$ \\
$2454732.0024$ &           &           & $ -45.76$ & $  91.06$ & $ -44.87$ & $  92.26$ \\
\enddata
\tablenotetext{a}{Radial velocities are reported only for epochs where
  there was a sufficient signal-to-noise ratio to obtain a usable
cross-correlation.  This was satisfied for all the epochs in the
emission line analysis but only for $21$ epochs in the absorption line
analysis, due to the higher signal-to-noise of the strong H$\alpha$
lines compared to the continuum.}
\end{deluxetable*}

Spectra were reduced using standard procedures in {\sc
  IRAF}\footnote{IRAF is distributed by the National Optical 
Astronomy Observatories, which are operated by the Association of
Universities for Research in Astronomy, Inc., under cooperative
agreement with the National Science Foundation.} \citep{tody1993} from
  the {\sc echelle} package to extract the spectra from the target
fiber and sky fiber simultaneously.  The TRES CCD is read out using
two amplifiers, which necessitates combining them before tracing and
extraction.  This was done using the {\sc mscred} package in IRAF,
applying a multiplicative gain correction to equalize the difference
in gain between the two readout electronics chains.  We then divided
by a normalized flat field exposure to correct for fringing in the
red-most regions of the spectrum, which has a significant effect on
the orders we used for the radial velocity analysis.

At the time of our observations, the TRES detector suffered from a
very high energetic particle hit rate, the source of which is under
investigation.  The particle hits were removed by median-combining
each set of multiple exposures for the flat fields, target and
calibrations, and then using the statistics of the median frame to
apply an upper envelope clip to the target frames themselves.  This
allowed us to avoid having to combine our target exposures to
eliminate the particle hits, which would smear out the radial velocity
variations slightly as a result of the extremely short orbital period
of GJ~3236.

The stability of the TRES instrument allowed us to define the aperture
trace for each fiber using the high signal-to-noise flat fields, which
were then used to extract the target and calibration spectra,
employing the ``optimal'' (in a least squares sense) weighting scheme
of \citet{hewett1985}.  The spectra were wavelength-calibrated using
the standard {\sc ecidentify} procedure, which employs the (known) 
dispersion relation of the \'echelle to solve for a wavelength
solution over all orders of the spectrum simultaneously.  This is
necessary due to the relative paucity of sufficiently intense lines
from the ThAr lamp in each order of the spectrum at wavelengths $>
7500\ \angstrom$.

\subsection{Summary of system properties}
\label{ssp_sect}

Table \ref{photparams} summarizes the known system properties, from
the literature (principally the proper motion survey of
\citealt{ls2005} used to select the target stars for the MEarth
survey) and our own $V$ and MEarth photometry, converted to the
standard Johnson-Cousins system.

\begin{deluxetable}{lr}
\tabletypesize{\normalsize}
\tablecaption{\label{photparams} Summary of the photometric and
  astrometric properties of the GJ~3236 system.}
\tablecolumns{2}
\tablewidth{3.2in}

\tablehead{
\colhead{Parameter} & \colhead{Value}
}

\startdata
$\alpha_{2000}$\tablenotemark{a,b}    & $03^h37^m14^s.08$ \\
$\delta_{2000}$\tablenotemark{a,b}    & $+69^\circ10\arcmin49\farcs8$ \\
$\mu_\alpha \cos \delta$\tablenotemark{b}       & $ 0\farcs142\ {\rm yr^{-1}}$ \\
$\mu_\delta$\tablenotemark{b}       & $-0\farcs132\ {\rm yr^{-1}}$ \\
\\
MEarth\tablenotemark{c}                  &$11.05 \pm 0.05$ \\
\\
$V_{\rm J}$\tablenotemark{c}             &$14.28 \pm 0.05$ \\
$I_{\rm C}$\tablenotemark{c}             &$11.55 \pm 0.05$ \\
\\
$J_{\rm 2MASS}$\tablenotemark{d}  & $9.806 \pm 0.023$ \\
$H_{\rm 2MASS}$\tablenotemark{d}  & $9.198 \pm 0.028$ \\
$K_{\rm 2MASS}$\tablenotemark{d}  & $8.967 \pm 0.022$ \\
\enddata

\tablenotetext{a}{Equinox J2000.0, epoch 2000.0.}
\tablenotetext{b}{From \citet{ls2005}.}
\tablenotetext{c}{Median of the measured out-of-eclipse fluxes from
  our light curves.  We estimate that the observational errors are
  dominated by the uncertainty in the photometric calibration, which
  is approximately $0.05\ {\rm mag}$ for both passbands.  $I_C$ was
  computed from the measured ${\rm MEarth}$ and $V_J$ magnitudes using
  Eq. (\ref{color_eq}).}
\tablenotetext{d}{We quote the combined uncertainties from the 2MASS
  catalog, noting that the intrinsic variability of our target means
  in practice that these are underestimates.}

\end{deluxetable}

The average color of $V_{\rm J} - I_{\rm C} = 2.73 \pm 0.07$
indicates an average spectral type of approximately M4 using the
color to spectral type relation of \citet{leggett1992} for young
Galactic disk stars.

\section{Initial light curve analysis ({\sc ebop})}
\label{ebop_lc_sect}

For detached eclipsing binaries with circular orbits, the radial
velocity (RV) and light curve models are largely independent.  We
therefore carried out a preliminary analysis of the available light
curves before starting to obtain spectroscopic observations.  The
principal purpose of doing this was to determine an extremely precise
orbital period to better-target the radial velocity observations, but
we can also constrain the system eccentricity by using the phase of
the secondary  eclipses relative to the primary eclipses (related to
$e \cos \omega$).  A simple geometric argument can then be used to
obtain a limit on the eccentricity itself, assuming no {\it a priori}
knowledge of $\omega$.

We used {\sc jktebop} (\citealt*{south2004a}; \citealt{south2004b}), a
modified version of {\sc ebop} (Eclipsing Binary Orbit Program;
\citealt{pe1981}; \citealt{e1980}), to obtain these parameters by
fitting the MEarth light curve.  The program was modified to fit
simultaneously for the EB model, and a simple form for the
synchronized out-of-eclipse modulations, assuming they can be
approximated by a truncated Fourier series:
\begin{equation}
m(t) = a_1 \sin(\Omega t) + b_1 \cos(\Omega t) + a_2 \sin(2 \Omega t)
+ b_2 \cos(2 \Omega t)
\label{ooe_eq}
\end{equation}
where $\Omega = 2 \pi / P$, and $P$ is the orbital period of the
binary.  The normalization term is omitted since this is already taken
into account by the standard EBOP model.  The revised code yields a
fit with very small residuals for the present system, and the use of the
form in Eq. (\ref{ooe_eq}) keeps the number of parameters required to
describe the modulations to a minimum compared to a full spot model.
Doing so will not necessarily reproduce the correct eclipse depths for
a given set of physical parameters (or vice versa), but should be
adequate for deriving an accurate system ephemeris.  We relax this
assumption in \S \ref{wd_sect}, where we adopt a full spot model for
determining the geometric and radiative parameters of the system.

The following parameters were allowed to vary in the fit: the sum of
the radii divided by the semimajor axis, $r_1 + r_2$ (where $r_j = R_j
/ a$, $R_j$ is the stellar radius of star $j$ and $a$ is the semimajor
axis), orbital inclination $i$, $e \cos \omega$, central surface
brightness ratio $J$ (essentially equal to the ratio of the primary
and secondary eclipse depths), orbital period $P$, orbital phase
zero-point $t_0$, light curve normalization, and 
the parameters $a_1$, $a_2$, $b_1$ and $b_2$ describing the out of
eclipse modulations.  Single-band light curves showing grazing
eclipses constrain the ratio of the radii only 
very weakly (or equivalently, the luminosity ratio), so these
parameters and the mass ratio (used for computing the oblateness of
the stars and the reflection effect) were fixed at $1.0$ for the
initial analysis.  The results for the orbital parameters were found
not to change significantly upon revising these to the measured values
from the spectroscopy once they were available.

Our light curves are not of sufficient quality to fit for the stellar
limb darkening, so we assumed a square root limb-darkening law with
coefficients appropriate to the $I$ (Cousins) passband from
\citet{claret2000} using the {\sc phoenix} model atmospheres.  Given
the lack of a spectroscopic constraint on the effective temperature,
surface gravity and metallicity for either star, we have instead
assumed solar metallicity, and derived temperatures and surface
gravities of $T_1 = 3280\ {\rm K}$ and $\log g_1 = 4.9$, and for the
secondary, $T_2 = 3200\ {\rm K}$ and $\log g_2 = 5.0$ by iterating the
modeling process.  The assumed temperatures were derived using the
masses from the combined orbital and light curve solution,
interpolating between the compilation of values for field stars by
\citet{leggett1992}.  See also \S \ref{kin_sect} where these
temperatures are verified using the final model.  A gravity darkening
exponent of $\beta = 0.32$, a value appropriate for stars with
convective envelopes \citep{lucy1967}, was also assumed, and the
option in {\sc jktebop} to calculate the reflection effect was used
rather than fitting for it.

We report only the orbital parameters ($P$, $t_0$ and $e \cos \omega$)
from the {\sc ebop} analysis, and have adopted these for the remainder
of this work.

The parameter uncertainties were derived using a Monte Carlo algorithm
built-in to {\sc jktebop} \citep{south2005}.  Briefly, this algorithm
uses the best fit to generate a synthetic light curve, injecting
Gaussian noise with amplitude determined by the observational errors
(which were scaled such that the reduced $\chi^2$ of the fit was equal
to unity) to produce a simulated light curve, which is then fit to
determine a new set of parameters.  The parameter uncertainties can be
estimated using the distribution of the parameters from a large number
of realizations (here, $10\,000$) of this process.  See also
\citet{south2004a,south2004b} for more details.

Times of minimum light derived for the $11$ well-sampled eclipses in
our $V$ and MEarth observations are reported in Table \ref{minlight}.
These were estimated using the method of \citet{kwee1956} over a $\pm
0.03$ region in normalized orbital phase around each eclipse.  We
subtracted the truncated Fourier series in Eq. (\ref{ooe_eq}) from the
light curves before applying this analysis to reduce the effect of any
systematic errors arising from the out-of-eclipse modulation.

\begin{deluxetable}{ll}
\tabletypesize{\normalsize}
\tablecaption{\label{orbparams} Derived orbital parameters for the GJ~3236 system.}
\tablecolumns{2}
\tablewidth{3.2in}

\tablehead{
\colhead{Parameter} & \colhead{Value\tablenotemark{a}}
}

\startdata
$P$       & $0.7712600 \pm 0.0000023\ {\rm days}$ \\
$t_0$     & $2454734.99586 \pm 0.00012\ {\rm HJD}$\tablenotemark{b} \\
\\
$e \cos \omega$ & $0.00008 \pm 0.00020$ \\
$e$            & $< 0.0022$ (95\% CL)\tablenotemark{c} \\
               & $< 0.0112$ (99\% CL)\tablenotemark{c} \\
\\
$\alpha$       & $0.60 \pm 0.04$ \\
\\
\hline
\multicolumn{2}{l}{Absorption line solution (adopted)}\\
\hline
\\
$\gamma$  & $10.06 \pm 0.94\ \kms$ \\
$K_1$     & $ 85.6 \pm 2.1\ \kms$ \\
$K_2$     & $114.8 \pm 1.9\ \kms$ \\
\\
$q$        & $0.746 \pm 0.023$ \\
$a \sin i$     & $3.053 \pm 0.044\ \rsun$ \\
$M_1 \sin^3 i$ & $0.368 \pm 0.015\ \msun$ \\
$M_2 \sin^3 i$ & $0.275 \pm 0.014\ \msun$ \\
\\
$\sigma_1$ &$6.3\ \kms$\tablenotemark{d} \\
$\sigma_2$ &$5.6\ \kms$ \\
\\
\hline
\multicolumn{2}{l}{H$\alpha$ emission line solution\tablenotemark{e}: GJ~856A}\\
\hline
\\
$\gamma$  & $12.87 \pm 0.19\ \kms$ \\
$K_1$     & $ 88.48 \pm 0.33\ \kms$ \\
$K_2$     & $114.71 \pm 0.49\ \kms$ \\
\\
$q$       & $0.7713 \pm 0.0045$ \\
$a \sin i$     & $3.0962 \pm 0.0091\ \rsun$ \\
$M_1 \sin^3 i$ & $0.3786 \pm 0.0037\ \msun$ \\
$M_2 \sin^3 i$ & $0.2919 \pm 0.0025\ \msun$ \\
\\
$\sigma_1$ &$1.3\ \kms$ \\
$\sigma_2$ &$1.9\ \kms$ \\
\\
\hline
\multicolumn{2}{l}{H$\alpha$ emission line solution\tablenotemark{e}: GJ~3379}\\
\hline
\\
$\gamma$  & $13.98 \pm 0.20\ \kms$ \\
$K_1$     & $ 88.65 \pm 0.34\ \kms$ \\
$K_2$     & $114.87 \pm 0.52\ \kms$ \\
\\
$q$       & $0.7717 \pm 0.0047$ \\
$a \sin i$     & $3.1011 \pm 0.0096\ \rsun$ \\
$M_1 \sin^3 i$ & $0.3802 \pm 0.0039\ \msun$ \\
$M_2 \sin^3 i$ & $0.2934 \pm 0.0026\ \msun$ \\
\\
$\sigma_1$ &$1.3\ \kms$ \\
$\sigma_2$ &$2.0\ \kms$ \\
\enddata

\tablenotetext{a}{We report $68.3\%$ confidence intervals, with error
  bars symmetrized by adopting the mean of the $15.85\%$ and $85.15\%$
  levels as the central value.}
\tablenotetext{b}{Ephemeris zero point, chosen to correspond to the
  epoch of the first primary eclipse in the $V$-band data from \S
  \ref{v_sect}.}
\tablenotetext{c}{Derived assuming only the measured $e \cos \omega$
  and a uniform distribution in $\omega$, for $95\%$ and $99\%$
  confidence levels.  These confidence levels are in fact lower
  limits, since there were no noticeable differences in the eclipse
  durations, or deviations from a circular orbit in radial velocity.
  $e = 0$ was assumed henceforth for the radial velocity modeling.}
\tablenotetext{d}{RMS of the residuals after subtracting the Keplerian
  orbit model from the data.  These are representative of the typical
  uncertainty on each RV point.}
\tablenotetext{e}{We re-emphasize that the H$\alpha$ radial
  velocities, despite having smaller random errors than the absorption
  line velocities, may have an unknown and potentially significant
  systematic error, due to the uncertainty in the source of the
  H$\alpha$ emission line, and whether it tracks the stellar
  photosphere.  We therefore conservatively adopt the absorption line
  solution for the remainder of the present publication.}

\end{deluxetable}

\begin{deluxetable}{lrrrl}
\tabletypesize{\normalsize}
\tablecaption{\label{minlight} Measured times of minimum light for GJ~3236.}
\tablecolumns{5}
\tablewidth{3.2in}

\tablehead{
\colhead{HJD} & \colhead{$(O - C)$ (s)} &
\colhead{Cycle\tablenotemark{a}} & \colhead{Band}
}

\startdata
$2454734.995569$ &$-32.5 \pm 52.0$ &$0.0$ &$V$ \\
$2454736.923700$ &$-34.5 \pm 26.7$ &$2.5$ &$V$ \\
$2454741.937128$ &$ -7.8 \pm 17.0$ &$9.0$ &MEarth \\
$2454742.708283$ &$-19.6 \pm 18.9$ &$10.0$ &MEarth\\
$2454745.793562$ &$  7.2 \pm 18.9$ &$14.0$ &MEarth\\
$2454762.761020$ &$-22.0 \pm 25.2$ &$36.0$ &MEarth\\
$2454765.846327$ &$  7.9 \pm 78.0$ &$40.0$ &MEarth\\
$2454766.617531$ &$  1.7 \pm 13.7$ &$41.0$ &MEarth\\
$2454767.774152$ &$-28.4 \pm 67.0$ &$42.5$ &MEarth\\
$2454770.859614$ &$ 18.9 \pm 36.6$ &$46.5$ &MEarth\\
$2454771.630638$ &$ -7.6 \pm 38.6$ &$47.5$ &MEarth\\
\enddata

\tablenotetext{a}{Cycle number, counting from $0$ at the primary
  eclipse at $t_0$, in units of the orbital period.  Integer numbers
  correspond to primary eclipses.}

\end{deluxetable}

\section{Radial velocity analysis}
\label{rv_sect}

Radial velocities were obtained using the two-dimensional
cross-correlation algorithm {\sc todcor} \citep{zm1994}, which uses
templates matched to each component of a spectroscopic binary to
simultaneously derive the velocities of both stars, and importantly
for the present application, an estimate of their luminosity ratio
($\alpha$) in the spectral bandpass.

We performed both a standard absorption line based cross-correlation
analysis, and an emission line analysis using the strong H$\alpha$
emission observed in both components of GJ~3236.  The signal to noise
ratio of the latter is substantially higher in our data, but we
caution that the source of the H$\alpha$ emission in our target stars
is not well-understood.  The emission line analysis results could
suffer systematic errors depending on the relative velocities of the
regions of the chromosphere emitting the H$\alpha$ and the stellar
photospheres, and this is exacerbated in the present system because it
is highly likely that the stellar spin and binary orbit are
synchronized, both from the light curves and expectations from tidal
theory \citep{zahn1977}.  We therefore conservatively adopt the
results from the absorption line analysis despite the larger errors.

The absorption line analysis used as a template spectrum a single
epoch observation of Barnard's star (\citealt{barnard1916}; also known
as GJ~699) taken on UT 2008 October 20.  We used a wavelength range of
8700 to 8850 {\AA} in order 50 of the spectrum to derive the
velocities, since this region contains a number of reasonably strong
metallic lines in M-dwarfs, and is free of telluric absorption lines.
The Ca\ {\sc II} infra-red triplet at 8498, 8542, and 8662 {\AA} was
deliberately avoided since the absorption in these lines is saturated,
and they are therefore intrinsically very broad, and emission cores
are often observed in these lines in active stars, which would
severely complicate the use of an inactive field star as a template.
We assumed a barycentric radial velocity of $-110.13\ \kms$ for
Barnard's star, derived from presently unpublished CfA Digital
Speedometer measurements spanning $17\ {\rm years}$.

For the emission line analysis, we used as templates spectra of two M4
dwarfs that were found to display H$\alpha$ emission, GJ~856A and
GJ~3379, observed as part of another program using the same
instrument.  GJ~856A was observed on UT 2008 October 17 and GJ~3379
was observed on UT 2008 October 19.  We used a wavelength range of
6520 to 6645 {\AA} in order 37 for the emission line
cross-correlations, noting that the cross-correlations are dominated
by the H$\alpha$ lines.

Barycentric radial velocities of GJ~856A and GJ~3379 were determined
by cross-correlation with Barnard's star, using the 8700 to 8850 {\AA}
region as for our absorption line analysis of GJ~3236.

Given that {\sc todcor} provides a measure of the correlation between
the observed and template spectra, it is possible to run
cross-correlations against a series of templates and use the
correlation value as an indicator of the template match.  Both the 
effective temperature and the rotational velocity of the templates
affect the solution, so it is vital to explore this parameter space.
Due to the availability of a limited range of observed templates, the
effective temperature could not be varied, but we were able to rotationally
broaden the template spectra in order to create primary and secondary
templates with $v_{\rm rot} \in \{1,2,4,6,8,10,12,16,20,25,30,35,40\}\
\kms$.  We used linear limb darkening coefficients $u_1 = 0.5782$ and
$u_2 = 0.5934$ from \citet{claret2004} for the SDSS $z$ passband (a
close match to the effective wavelength of the spectroscopy), which
correspond to our assumed temperatures for the primary and secondary
(see \S \ref{ebop_lc_sect}).

We ran {\sc todcor} on the grid of $169$ template combinations, and
sought the maximum of the resulting correlation values to determine
the most likely $v_{\rm rot} \sin i$ values for the two components of
the binary system.  Barnard's star is a very slow rotator, so the
rotation of the template is negligible given our spectral resolution.
Our analysis indicates $v_{\rm rot,1} \sin i = 25\ \kms$, which is
consistent with the expected value of $v_{\rm rot,1} \sin i = 24.6\
\kms$ if the spin and orbit are synchronized.  $v_{\rm rot,2} \sin i$
was poorly-constrained, with the maximum correlation occurring at $3\
\kms$, whereas we would expect $19.5\ \kms$ assuming synchronization.
We have simply adopted the latter value, since the light curves and
primary velocity indicate that the system is indeed synchronized.
Using the templates with $v_{\rm rot,1} \sin i = 25\ \kms$ and $v_{\rm
  rot,2} \sin i = 20\ \kms$, {\sc todcor} indicates $\alpha = 0.60 \pm
0.04$.

For the emission line analysis, we assumed the stellar rotation values
from the absorption line analysis, rotationally broadening the
observed spectra of GJ~856A and GJ~3379.  The intrinsic rotational
broadening in both of these templates is again negligible for our
purposes.  {\sc todcor} gave a maximum correlation for $\alpha_{\rm
  Ha} = 0.52 \pm 0.02$, which provides a measure of the relative
H$\alpha$ emission luminosities of the two stars.

\begin{figure}
\centering
\includegraphics[angle=0,width=2.8in]{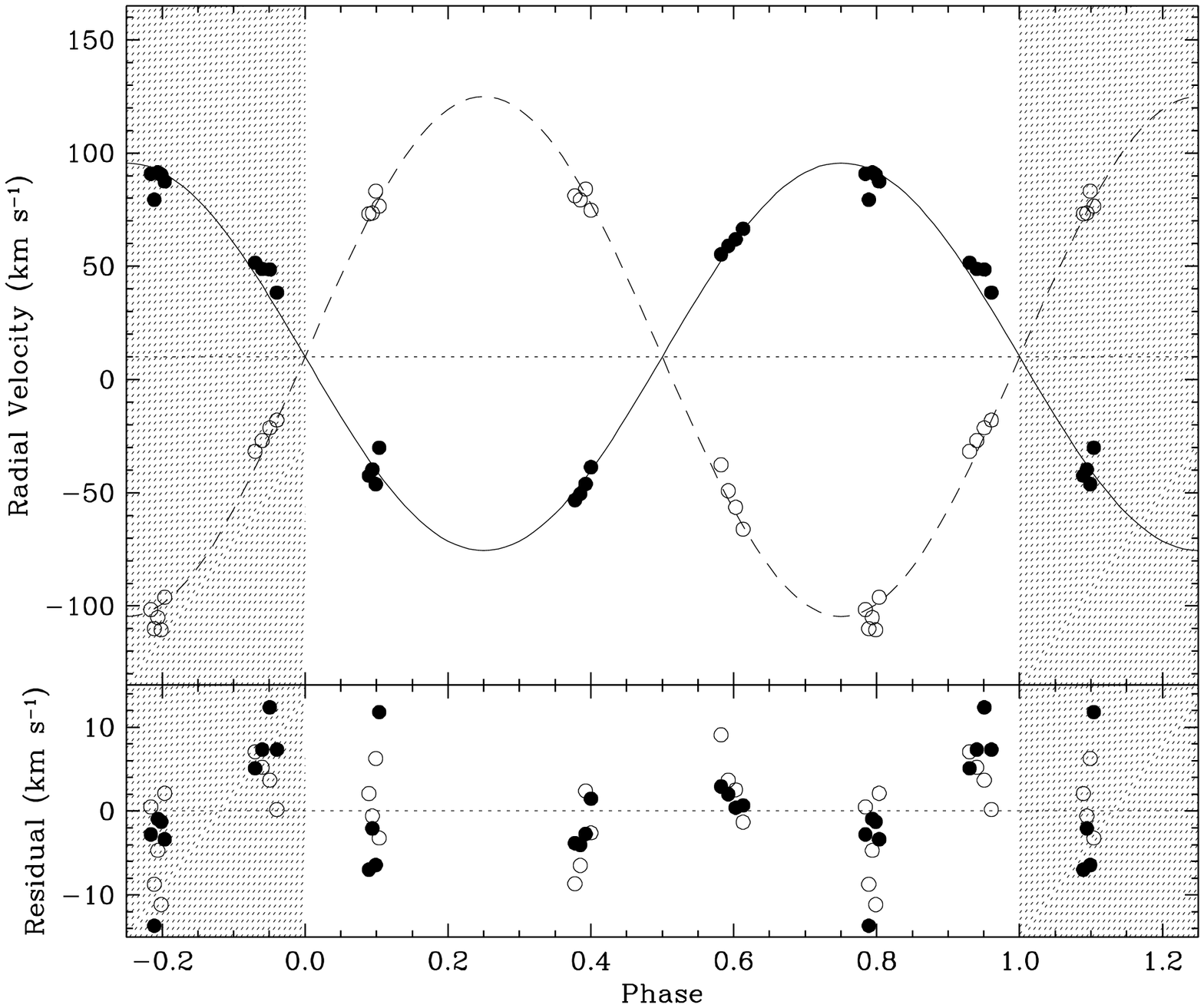}

\includegraphics[angle=0,width=2.8in]{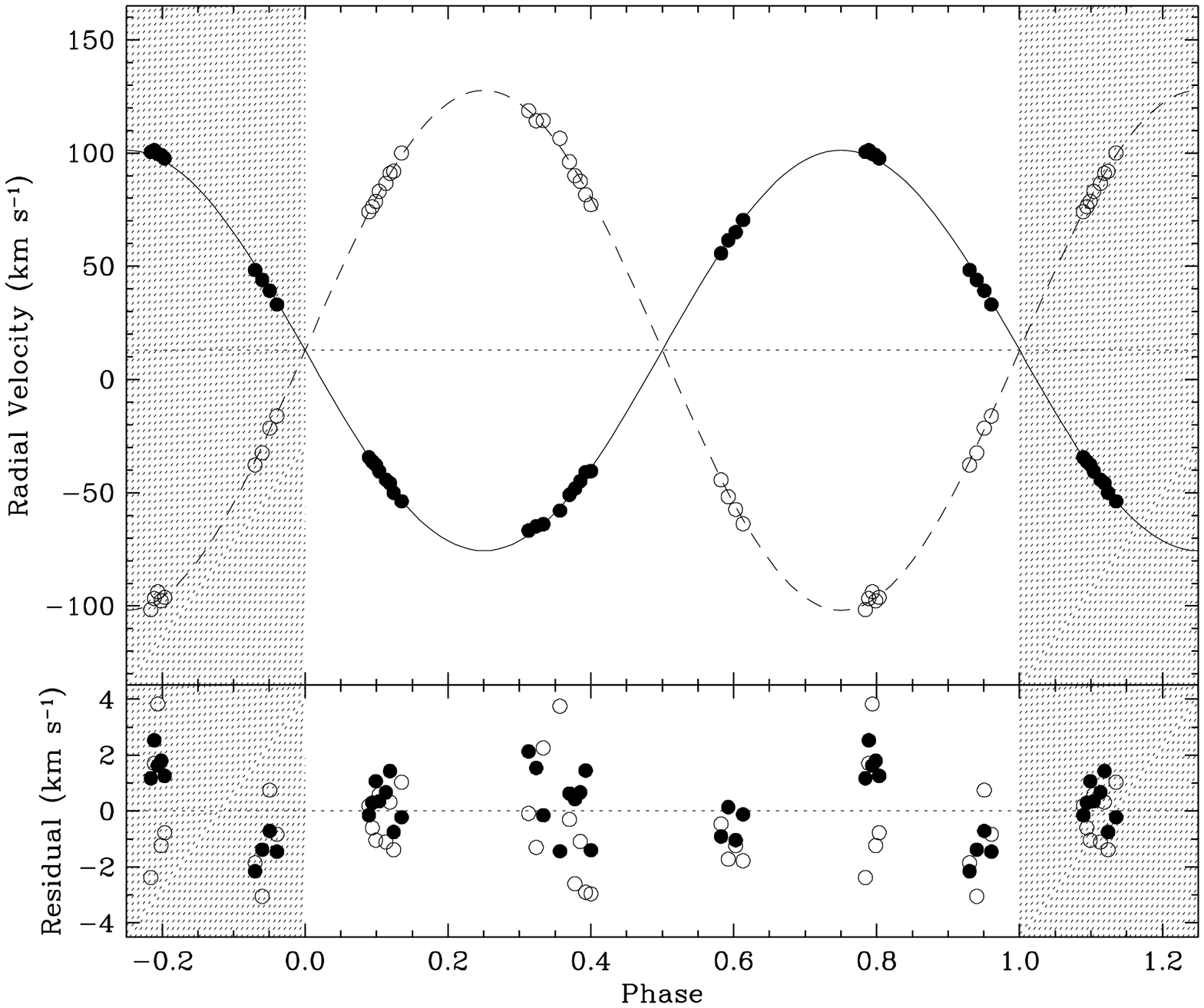}

\includegraphics[angle=0,width=2.8in]{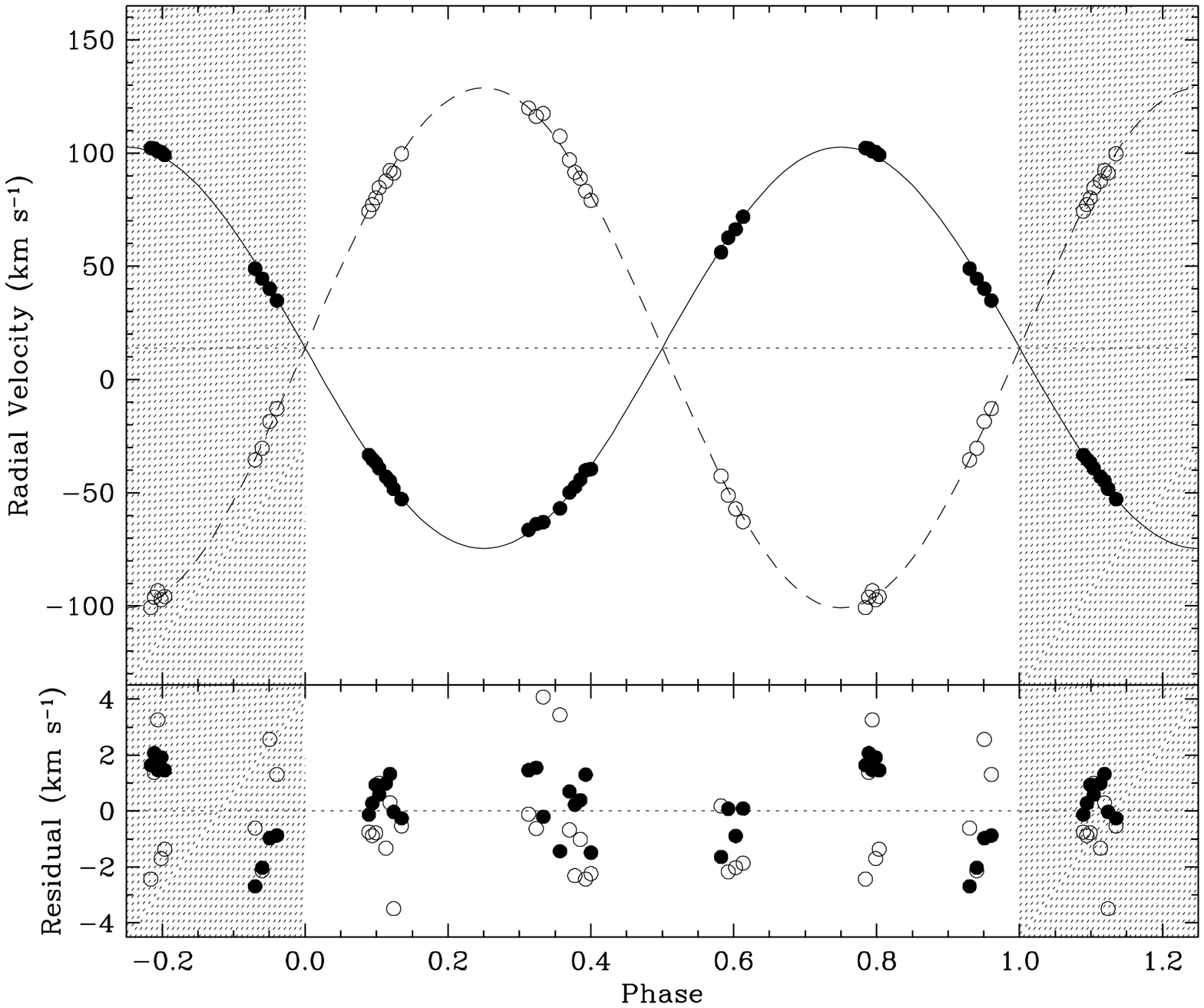}
\caption{Phase-folded radial velocity curves for the GJ~3236 system.
  Shown in each panel are the radial velocities for the primary
  (filled symbols) and the secondary (open symbols), with the
  best-fitting Keplerian orbit overplotted using solid and dashed
  lines for the two components, respectively.  The lower sub-panels
  show the residuals after subtracting the model from the data.  Top:
  absorption line solution cross-correlating against Barnard's star;
  Center: emission line solution using GJ~856A; Bottom: emission line
  solution using GJ~3379 as the template.}

\label{rv_fit}
\end{figure}

Since the orbital period and time of primary eclipse are extremely 
well-determined from the light curves, we held these values fixed in
the final radial velocity orbital solution.  We also assumed a circular
orbit.  Figure \ref{rv_fit} shows the resulting radial velocity curves.
The parameters derived from the radial velocity analysis are given in
Table \ref{orbparams}.  We assumed equal observational errors on each
radial velocity point, scaled such that the reduced $\chi^2$ of the
fit was equal to unity.

Radial velocities of the components of double-lined spectroscopic
binaries are prone to a ``peak pulling'' effect as they approach the
$\gamma$ velocity (at times of minimum separation of the two
components in the spectrum).  The {\sc todcor} method can largely
eliminate this, provided the template spectra match the target
sufficiently well.  In our experience, this effect is particularly
sensitive to errors in the assumed rotational broadening: if the
template contains too much or too little rotation, the velocities
would be pushed toward or away from $\gamma$ to compensate.  Examining
the residuals of our solution in Figure \ref{rv_fit} indicates little
or no ``peak pulling'', which further justifies our assumed rotational
velocities and the assumption of negligible rotation in the templates.

\section{Light curve analysis (Wilson \& Devinney code)}
\label{wd_sect}

In order to properly account for spots on the components of GJ~3236,
we use the most recent version (WD2007) of the standard \citet{wd1971}
code to derive the geometric and radiative parameters of the system.
This code incorporates a full physical model including spots and using
model atmospheres.

WD normally uses the \citet{kurucz1979,kurucz1993} model atmospheres
to calculate the emergent flux from surface elements on each binary
component.  These models span an effective temperature range of
$3500-50000\ \kelvin$, and for temperatures below $3500\ \kelvin$, WD
interpolates between a black body of the appropriate temperature and 
the $3500\ \kelvin$ model atmosphere.  This is weighted in such a
fashion as to produce a pure black body at $1500\ \kelvin$, and for
intermediate temperatures there is a smooth ramp toward a pure model
atmosphere at $3500\ \kelvin$.  Molecular sources of opacity
increasingly begin to dominate the optical spectra of M-dwarfs in this
temperature range, so a black body is clearly a poor approximation for
the emergent flux.  Since both components of GJ~3236 appear to be
cooler than $3500\ \kelvin$, we have modified our copy of WD to
substitute {\sc phoenix} NextGen model atmospheres
\citep*{hauschildt1999} in place of these interpolations for effective
temperatures cooler than $3500\ \kelvin$.

\subsection{Assumptions}
\label{wd_assump_sect}

Our assumptions for the WD light curve model are detailed in Table
\ref{wd_pars}.  We have assumed zero orbital eccentricity,
synchronized spin and orbit for both components of the binary, and no
third light, which are justified in \S \ref{mearth_obs_sect} and \S
\ref{ebop_lc_sect}.

\begin{deluxetable*}{llll}
\tabletypesize{\normalsize}
\tablecaption{\label{wd_pars} Parameters used in the WD light curve model.}
\tablecolumns{4}

\tablehead{
\colhead{Parameter} & \colhead{WD name} & \colhead{Value or
  ``varied''} & \colhead{Description}
}

\startdata
\multicolumn{2}{l}{Binary type} &detached \\
\\
$t_0$   &{\tt HJD0}   &$2454734.99586$ (HJD) &Phase zero-point\tablenotemark{a} \\
$P$     &{\tt PERIOD} &$0.77126\ {\rm days}$  &Orbital period\tablenotemark{a} \\
$dP/dt$ &{\tt DPDT}   &$0$                    &First derivative of period \\
\\
$a \sin i$ &{\tt SMA} $\times \sin i$ &$3.053 \pm 0.044\ \rsun$ &Projected semimajor axis\tablenotemark{a,b} \\
$q$        &{\tt RM}  &$0.746 \pm 0.023$      &Mass ratio\tablenotemark{a} \\
$\gamma$   &{\tt VGA} &$10.06 \pm 0.94\ \kms$ &Systemic velocity\tablenotemark{a} \\
\\
$e$     &{\tt E}      &$0$                    &Orbital eccentricity \\
$F_1$   &{\tt F1}     &$1.0$                  &Primary rotation parameter \\
$F_2$   &{\tt F2}     &$1.0$                  &Secondary rotation parameter \\
\\
        &{\tt HLA}    &varied                 &Light curve normalization \\
\\
$T_1$   &{\tt TAVH}   &$3280\ {\rm K}$        &Primary effective temperature \\
$T_2$   &{\tt TAVC}   &varied                 &Secondary effective temperature \\
$\Omega_1$ &{\tt PHSV}  &varied              &Primary surface potential\tablenotemark{c}  \\
$\Omega_2$ &{\tt PCSV}  &varied              &Secondary surface potential\tablenotemark{c} \\
\\
$i$     &{\tt INCL}   &varied                 &Orbital inclination \\
$L_3$   &{\tt EL3}    &$0$                    &Third light \\
\\
$A_1$   &{\tt ALB1}   &$0.5$ &Primary bolometric albedo\tablenotemark{d} \\
$A_2$   &{\tt ALB2}   &$0.5$                  &Secondary bolometric albedo\tablenotemark{d} \\
$\beta_1$   &{\tt GR1}    &$0.32$            &Primary gravity darkening exponent\tablenotemark{d} \\
$\beta_2$   &{\tt GR2}    &$0.32$            &Secondary gravity darkening exponent\tablenotemark{d} \\
\\
$\theta_{s,1}$ &{\tt XLAT1} &$60^\circ$           &Spot 1 latitude\tablenotemark{e} \\
$\phi_{s,1}$   &{\tt XLONG1} &varied          &Spot 1 longitude \\
$r_{s,1}$      &{\tt RADSP1} &varied          &Spot 1 radius \\
$T_{s,1}/T_1$  &{\tt TEMSP1} &0.96            &Spot 1 temperature contrast \\
\\
$\theta_{s,2}$ &{\tt XLAT2} &$60^\circ$           &Spot 2 latitude\tablenotemark{e} \\
$\phi_{s,2}$   &{\tt XLONG2} &varied          &Spot 2 longitude \\
$r_{s,2}$      &{\tt RADSP2} &varied          &Spot 2 radius \\
$T_{s,2}/T_2$  &{\tt TEMSP2} &0.96            &Spot 2 temperature contrast \\
\\
$\alpha$       &     &$0.60 \pm 0.04$         &Light ratio\tablenotemark{a} \\
\enddata

\tablenotetext{a}{From Table \ref{orbparams}}
\tablenotetext{b}{Although the WD parameter {\tt SMA} is the semimajor
  axis itself, we fixed $a \sin i$ in our MCMC procedure to the value
  from the radial velocities, calculating the required {\tt SMA} value
  using $i$ ({\tt INCL}) at each iteration.}
\tablenotetext{c}{Modified \citet{kopal1959} potentials.  These are
    roughly inversely proportional to stellar radii for this detached
    binary; for a clear description of their definition, we refer the
    reader to the documentation for the new WD graphical user
    interface {\sc phoebe} \citep{prsa2005}.}
\tablenotetext{d}{Values appropriate for convective atmospheres.}
\tablenotetext{e}{See text.}

\end{deluxetable*}

The limited phase coverage and lack of repeat observations in our
$V$-band photometry (leading to an uncertainty in any corrections for
color-dependent ``second-order'' atmospheric extinction), and poor
precision (large scatter) in the $J$-band photometry means in practice
that these are not useful in aiding the determination of the geometric
(and thus physical) parameters of the system.  Our experiments with
the WD model indicate that these provide minimal additional
constraints on any of the parameters when compared to using the MEarth
data alone.  We have therefore elected to use only the MEarth data for
fitting for the system parameters, and then evaluated this model with
respect to the other passbands as a check.

Doing so necessitates fixing one of the component effective
temperatures, since the relative eclipse depths measured in a single
passband merely determine the relative temperatures of the two
components.  We have therefore assumed $T_1 = 3280\ {\rm K}$ following
\S \ref{ebop_lc_sect}.  In practice even given multi-band photometry
it is usually necessary to fix this parameter: although the
color-dependence of the eclipse depths does in principle provide
information on the value of $T_1$, this is normally poorly constrained
due to the difficulty of measuring differential colors to high
precision.

In order to reproduce the out-of-eclipse variations seen in our
optical light curves, at least two spots must be introduced into the
model.  The configuration of these spots affects the measured eclipse
depths, depending on whether a spotted or unspotted part of the
photosphere is eclipsed.  This predominantly affects the derived
ratio of surface brightnesses as measured by the ratio of depths of
the primary and secondary eclipses (and thus the ratios of the
component radii and effective temperatures inferred from this), and a
combination of $(R_1+R_2)/a$ and the orbital inclination inferred from
the total eclipse depth.

Since our $V$-band data do not cover the portions of the light curve
out of eclipse, and the scatter in the $J$-band data is too large to
usefully constrain the model, the spot modeling must be done using
single-band light curve information.  Spots are introduced in the WD
code using a standard four-parameter model, where each spot is
parametrized using the spot latitude $\theta_s$ and longitude
$\phi_s$, radius $r_s$ and temperature contrast $T_s / T_p$.  The
latter quantity gives the ratio of effective temperatures of the
spotted and unspotted photosphere.  For a single-band light curve, the
spot longitude is well-constrained by the phase of the out of eclipse
modulation, but the combination of $\theta_s$, $r_s$ and $T_s / T_p$
is largely degenerate (these quantities determine the amplitude and
shape of the variation).  Moreover, we can place each spot on either
star. Although the data do rule out some of the possible spot
locations, there are still a range of possibilities, which introduces
a systematic error into the geometric parameters of the system.
In principle, measuring the color-dependence of the out-of-eclipse
modulation would allow this degeneracy to be reduced.

Given the presently available data, we instead probe this systematic
error by considering three characteristic spot configurations that are
consistent with the light curve.  These are: (1) one close to polar
spot on each star, with $\theta_s = 60^\circ$, with both spots located
in the same hemisphere as the chord traversed during the eclipse,
resulting in the spot on the primary being eclipsed by the secondary
during primary eclipse, and vice versa during secondary eclipse; (2)
likewise, only with the spots located in the opposite hemisphere,
resulting in no spots being eclipsed; and (3) a case intermediate
between these, with a ``symmetric'' spot configuration of two spots on
each star, at latitudes of $60^\circ$ with one located in each
hemisphere.  In each model, the spot latitudes and temperature ratios
were held fixed, allowing only longitude and radius to vary, and in
configuration (3) we enforced symmetry of the spot pattern about the
equator, i.e. both spots on one star were forced to have the same
longitudes and radii.  This means that all three models have the same
number of parameters ($4$).

\subsection{Model fitting and error estimates}

To derive robust error estimates, including the correlations between
parameters, we have implemented a Markov Chain Monte Carlo (MCMC)
algorithm around the WD light curve model.  Following
\citet{ford2005}, we used the Metropolis-Hastings algorithm
\citep{metropolis1953,hastings1970} to estimate the {\it a posteriori}
joint probability distribution of the fitted model parameters.  The
fitting statistic was the traditional $\chi^{2}$, where the
observational error estimates were scaled such that each light curve's
reduced $\chi^2$ was equal to unity.

We briefly summarize the Metropolis-Hastings algorithm here.
Starting from an initial point in parameter space, the algorithm takes
the most recent set of parameters and perturbs one or more parameters
by a random Gaussian deviate.  If the perturbed parameter set has a
lower $\chi^2$ than its progenitor, it is accepted as a new point in
the chain.  If it has a larger $\chi^2$, it is accepted with a
probability $\exp(-\Delta \chi^2/2)$.  If it is not accepted, the
original point is repeated in the chain.  The size of the
perturbations were adjusted so that $20-30\%$ of the proposed points
were accepted.

We included the spectroscopic light ratio between the primary and
secondary components as an observation in our $\chi^2$ statistic,
which in practice was implemented by adding an extra contribution to
$\chi^2$ of the form:
\begin{equation}
\Delta \chi^2_{\rm prior} = \left[\frac{\alpha_{\rm measured} -
\alpha_{\rm WD}}{\sigma_\alpha}\right]^{2}
\end{equation}
where $\alpha_{\rm measured}$ and $\sigma_\alpha$ are the observed
light ratio and its error, taken from Table \ref{wd_pars}, and
$\alpha_{\rm WD}$ is the value calculated from the WD model.  This is
necessary for systems with near-circular orbits and grazing eclipses
because the light curves only very weakly constrain this quantity (or
equivalently, the ratio of the component radii).  Uninformative
(uniform) priors were assumed on all other parameters.

\begin{figure}
\centering
\includegraphics[angle=0,width=3.3in]{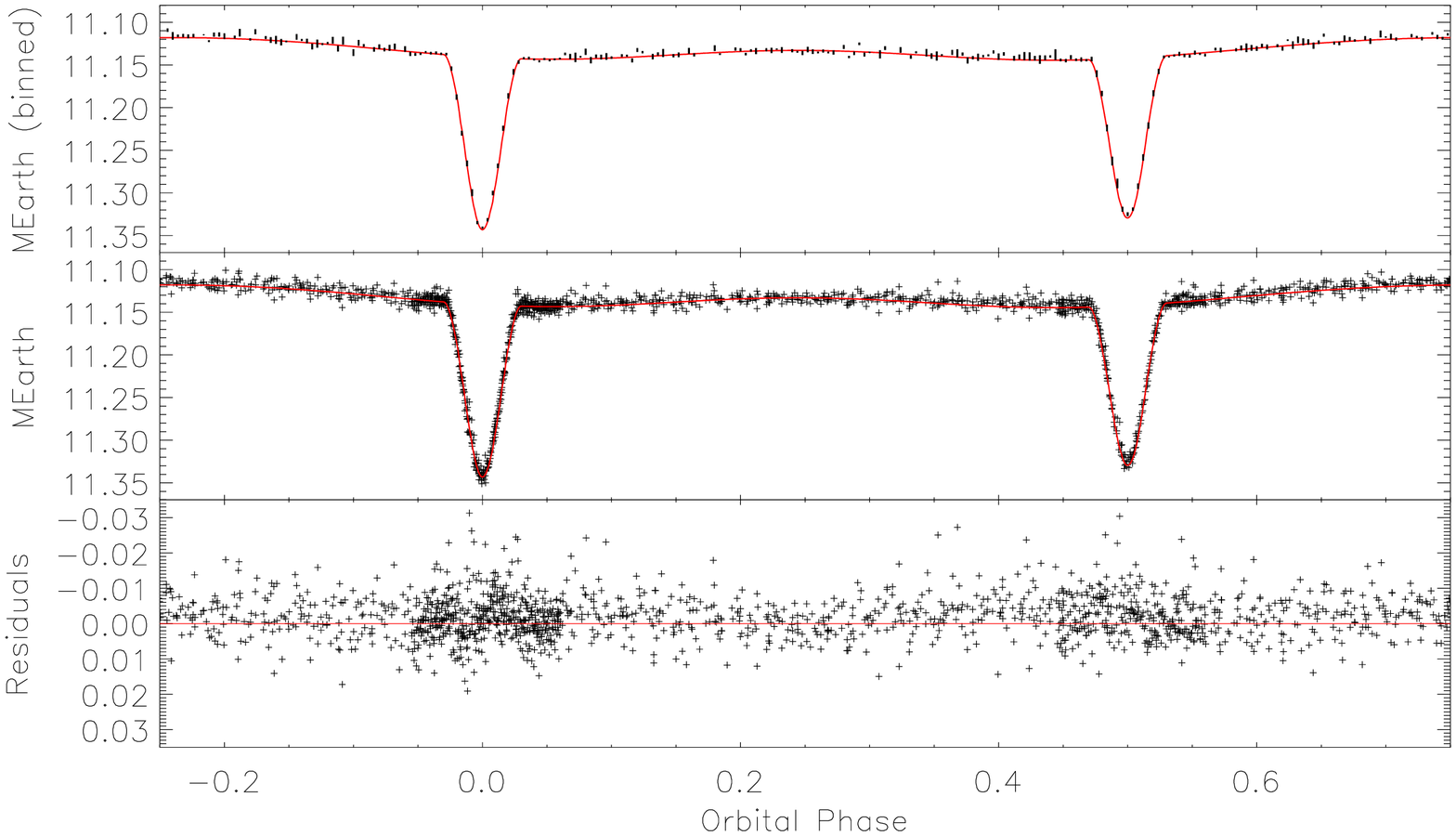}
\includegraphics[angle=0,width=3.3in]{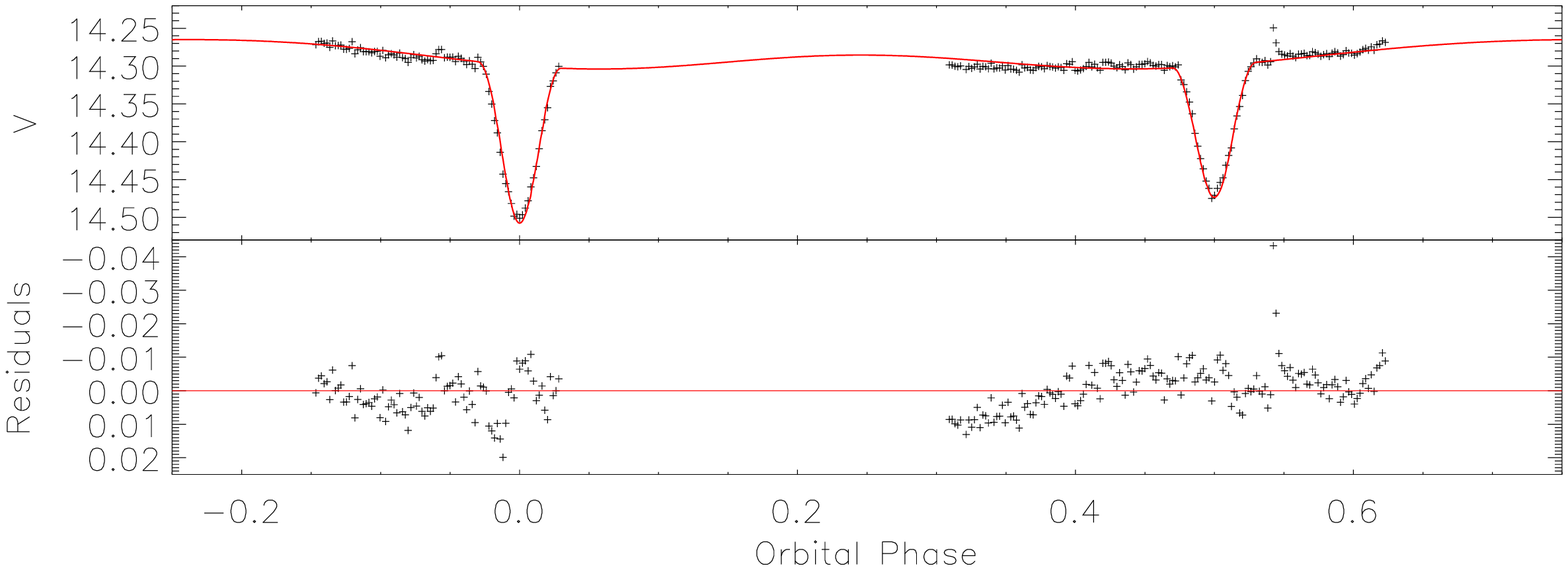}
\includegraphics[angle=0,width=3.3in]{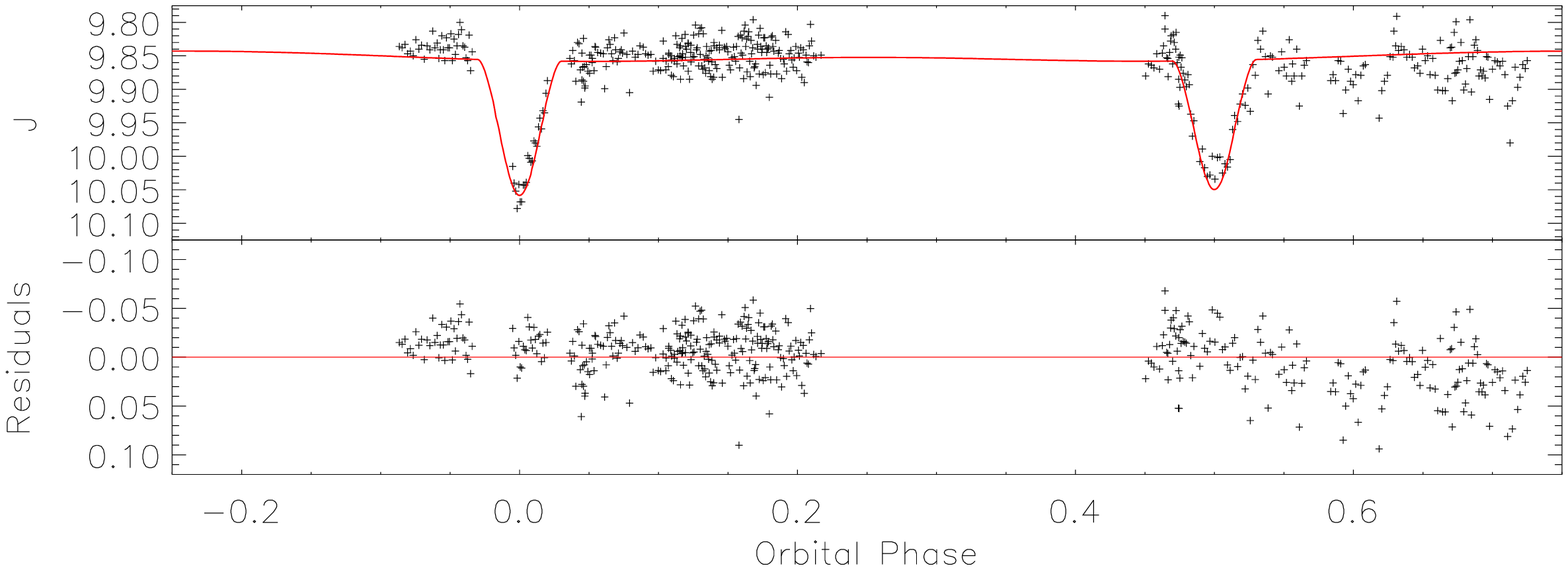}
\caption{Phase-folded light curves in MEarth, $V$ and $J$ passbands
  (top, middle and bottom panels, respectively).  In each panel, the
  upper sub-panels show the light curve (black points) with the WD
  model $3$ (see Table \ref{params}) overplotted (solid lines), and the
  lower sub-panels show the residuals after subtracting the model from
  the data.  The uppermost panel for the MEarth data shows a binned
  light curve (in 250 bins each spanning $0.004$ in normalized orbital
  phase) to allow the features to be more clearly distinguished given
  the large number of data points. }
\label{lc_model}
\end{figure}

The final parameters and error estimates were determined by taking the
$68.3\%$ confidence interval ($15.85$ and $84.15$ percentiles,
corresponding to $\pm 1$ standard deviation for a Gaussian
distribution) of the samples generated by our MCMC procedure.  The
chains were run until they contained at least $10^6$ samples, and we
discarded the first $10\%$ of the points in each chain in order
to ensure they had converged.  These parameters are reported in Table
\ref{params}, and Figure \ref{lc_model} shows the light curves with
the best-fitting model overplotted.

\begin{deluxetable*}{llll}
\tabletypesize{\normalsize}
\tablecaption{\label{params} Derived geometric, radiative and physical parameters of the GJ~3236 system.}
\tablecolumns{4}

\tablehead{
\colhead{Parameter} & \multicolumn{3}{c}{Value\tablenotemark{a}} \\
                    & \colhead{Model 1} & \colhead{Model 2} & \colhead{Model 3}
}

\startdata
{\tt HLA}    &$0.0002759 \pm 0.0000073$   &$0.0003037 \pm 0.0000014$ &$0.0002915 \pm 0.0000066$ \\
$T_2/T_1$    &$0.9801 \pm 0.0016$         &$0.97914 \pm 0.00085$ &$0.9774 \pm 0.0014$ \\
$\Omega_1$   &$9.01 \pm 0.12$             &$8.781 \pm 0.030$ &$8.794 \pm 0.098$ \\
$\Omega_2$   &$8.39 \pm 0.14$             &$9.236 \pm 0.041$ &$8.806 \pm 0.156$ \\
$i$          &$82.805 \pm 0.041$          &$83.721 \pm 0.032$ &$83.154^\circ \pm 0.047^\circ$ \\
\\
$\phi_{s,1}$ &$-17.2^\circ \pm 1.5^\circ$ &$-14.0^\circ \pm 1.4^\circ$ &$-21.6^\circ \pm 2.0^\circ$ \\
$r_{s,1}$    &$36.5^\circ \pm 1.1^\circ$  &$43.83^\circ \pm 0.41^\circ$ &$25.90^\circ \pm 0.42^\circ$ \\
$\phi_{s,2}$ &$17.5^\circ \pm 1.4^\circ$  &$20.4^\circ \pm 1.2^\circ$ &$22.0^\circ \pm 1.1^\circ$ \\
$r_{s,2}$    &$46.7^\circ \pm 1.8^\circ$  &$59.98^\circ \pm 0.93^\circ$ &$34.54^\circ \pm 0.65^\circ$ \\
\\
$(R_1+R_2)/a$ &$0.22397 \pm 0.00082$      &$0.21670 \pm 0.00073$  &$0.22137 \pm 0.00085$ \\
$R_2/R_1$     &$0.850 \pm 0.028$          &$0.7380 \pm 0.0048$    &$0.783 \pm 0.024$ \\
\\
$M_1$        &$0.377 \pm 0.016$           &$0.375 \pm 0.016$ &$0.376 \pm 0.017\ \msun$ \\
$M_2$        &$0.282 \pm 0.015$           &$0.280 \pm 0.015$ &$0.281 \pm 0.015\ \msun$ \\
$R_1$        &$0.3729 \pm 0.0078$         &$0.3829 \pm 0.0057$ &$0.3828 \pm 0.0072\ \rsun$ \\
$R_2$        &$0.3167 \pm 0.0075$         &$0.2828 \pm 0.0043$ &$0.2992 \pm 0.0075\ \rsun$ \\
\\
$a$          &$3.077 \pm 0.044$           &$3.071 \pm  0.044$  &$3.075 \pm 0.044\ \rsun$ \\
$\log g_1$   &$4.872 \pm 0.015$           &$4.8456 \pm 0.0081$ &$4.850 \pm 0.014$ \\
$\log g_2$   &$4.886 \pm 0.019$           &$4.9819 \pm 0.0115$ &$4.935 \pm 0.019$ \\
$v_{\rm rot,1}$ &$24.44 \pm 0.52$         &$25.12 \pm 0.37$    &$25.05 \pm 0.50$ \\
$v_{\rm rot,2}$ &$20.77 \pm 0.47$         &$18.54 \pm 0.27$    &$19.60 \pm 0.44$ \\
\\
$N_{\rm samp}$ &$1.61 \times 10^6$ &$2.35 \times 10^5$ &$1.37 \times 10^6$\\
$\sigma_{\rm MEarth}$\tablenotemark{b}  &$0.0061$  &$0.0061$  &$0.0061$ \\
$\sigma_V$\tablenotemark{b}             &$0.0074$  &$0.0076$  &$0.0066$ \\
$\sigma_J$\tablenotemark{b}             &$0.027$   &$0.028$   &$0.027$ \\
\enddata

\tablenotetext{a}{We report $68.3\%$ confidence intervals, with error
  bars symmetrized by adopting the mean of the $15.85\%$ and $85.15\%$
  levels as the central value.}
\tablenotetext{b}{RMS scatter of the residuals after subtracting the
  model from the data.}

\end{deluxetable*}

\subsection{Discussion}
\label{wd_disc_section}

The most serious limitations in the present analysis result from the
use of only single-band light curves to model the out of eclipse
variations.  This is clear from comparing the results for the three
solutions reported in Table \ref{params}, where we find significant
($> 1$ standard deviation) differences in the orbital inclination $i$,
ratio of effective temperatures $T_2/T_1$, and in $(R_1+R_2)/a$,
which depend on the assumptions made regarding the locations of the
spots on the components of the binary.

Given our radial velocity errors, the orbital inclination uncertainty
has little effect on the final physical parameters, as evidenced by
the component masses and semimajor axis reported in Table \ref{params},
which are essentially identical to within the observational errors.
The dominant uncertainty in the final solution is therefore in the
component radii.

Of the three models considered, the symmetric spot distribution in
model (3) seems the most physically realistic given the known
properties of spots on low-mass stars.  These are often found to be
polar in light curve models (e.g. \citealt{rodono1986}).  Moreover,
Doppler imaging studies (e.g. \citealt{barnes2000,barnes2001})
indicate that in reality, the surfaces of low-mass stars probably have
many small spots distributed over a range of latitudes.  This
situation is largely indistinguishable from a small number of large, polar
spots in one-dimensional light curve data, although it is important to
note that the effects on the measured eclipse depths {\it could} be
different.  There is no a-priori reason to expect the spots to be
concentrated into one hemisphere given the close to edge-on
inclination of the binary orbit.  We  therefore favor model (3), but
report all three solutions to provide a realistic evaluation of the
systematic error in our results, and reiterate that the data do not
distinguish between the three possibilities, with all of these having
comparable $\chi^2$ values and residuals.

As a check, we compare the model fit to the MEarth data alone with our
$V$ and $J$-band data in Figure \ref{lc_model}.  In order to do this,
we have refit the model for these bands allowing only the light curve
normalization parameter ({\tt HLA}) to vary, with all other parameters
fixed to the values determined from the MEarth data.  The model is
consistent with the $J$-band data within the scatter, and is
reasonably consistent with the $V$-band data, considering that the
early parts of the secondary eclipse curve were taken during quite
non-photometric conditions, and that we have not accounted for
color-dependent or differential atmospheric extinction effects, which
are expected to be larger in this band than the redder MEarth or
$J$ bandpasses, and cannot readily be determined due to the lack of
repeat observations.

The presence of spots and proximity effects in very close binary
systems have a small influence on the shape of the radial velocity
curve, causing it to depart from a Keplerian orbit as we assumed in \S
\ref{rv_sect}.  We have investigated the influence of these effects on
our results by comparing radial velocity curves generated using the WD
model and the simple Keplerian model, finding that the corrections to
radial velocities taken out of eclipse are dominated by the spots, the
properties of which are largely unknown.  However these are very small
($< 0.1\ \kms$) compared to the uncertainties in our radial velocity
measurements (approximately $6\ \kms$), so we are therefore justified
in neglecting them for the present analysis.

\section{Effective temperatures, luminosities and kinematics}
\label{kin_sect}

Provided the effective temperature of one component of an eclipsing
binary is known, the combined light curve and radial velocity
parameters are then sufficient to determine intrinsic, bolometric
luminosities of both components of the system.  This follows directly
from the definition of the effective temperature in terms of the
Stefan-Boltzmann law:
\begin{equation}
L_{\rm bol} = 4 \pi R^2 \sigma T^4
\end{equation}

By using bolometric corrections and the measured system magnitudes, we
can then infer the distance, provided the reddening can be constrained
or assumed to be zero.  In the present case, the latter is a
reasonable assumption, since our target is very nearby
(\citealt{gj1991} give a ``photometric parallax'' of $47 \pm 8\ {\rm
  pc}$, and \citealt{lepine2005} gives $21.3 \pm 4.4\ {\rm pc}$; note
that both of these assume the system is a single star, which means
they underpredict the distance to a near equal luminosity binary such
as the present one by a factor of approximately $\sqrt{2}$).

In order to determine the effective temperatures, we assume the
intrinsic colors and bolometric corrections for M-dwarfs compiled by
\citet{bessell1995,bessell1991} and \citet{bessell1988} to convert the
integrated system $V - I$ color presented in Table \ref{photparams}
into $T_1$, assuming the measured effective temperature ratio and
radii for both components from the light curve model.  We used the
polynomial fits of \citet{hillenbrand1997}, which cover the required
spectral type range, and assume an uncertainty of $0.1\ {\rm mag}$ in
these fits as stated in her Appendix C.  We also assume a $\pm 100\
{\rm K}$ systematic uncertainty in the effective temperature scale,
which has been included and propagated in the errors we report.
Table \ref{kinparams} gives our derived parameters for the GJ~3236
system.

\begin{deluxetable*}{llll}
\tabletypesize{\normalsize}
\tablecaption{\label{kinparams} Effective temperatures, luminosities
  and kinematics for GJ~3236.}
\tablecolumns{4}

\tablehead{
\colhead{Parameter} & \multicolumn{3}{c}{Value} \\
                    & \colhead{Model 1} & \colhead{Model 2} & \colhead{Model 3}
}

\startdata
$T_1$           & $3313 \pm 110\ {\rm K}$ & $3310 \pm 110\ {\rm K}$ & $3313 \pm 110\ {\rm K}$ \\
$T_2$           & $3247 \pm 108\ {\rm K}$ & $3241 \pm 108\ {\rm K}$ & $3238 \pm 108\ {\rm K}$ \\
$L_{\rm bol,1}$ & $0.0152 \pm 0.0021\ \lsun$ & $0.0160 \pm 0.0022\ \lsun$ & $0.0160 \pm 0.0021\ \lsun$ \\
$L_{\rm bol,2}$ & $0.0101 \pm 0.0014\ \lsun$ & $0.0080 \pm 0.0011\ \lsun$ & $0.0089 \pm 0.0012\ \lsun$ \\
\\
$M_V$        & $11.17 \pm 0.30$ & $11.23 \pm 0.30$ & $11.19 \pm 0.30$ \\
$M_I$        & $8.44 \pm 0.27$ & $8.50 \pm 0.27$ & $8.46 \pm 0.27$ \\
\\
$(m-M)$      & $3.11 \pm 0.28$ & $3.05 \pm 0.28$ & $3.09 \pm 0.28$ \\
$d$          & $42.2 \pm 5.5\ {\rm pc}$ & $41.1 \pm 5.3\ {\rm pc}$ & $41.8 \pm 5.4\ {\rm pc}$ \\
\\
$U$          & $+34.2 \pm 3.9\ \kms$ & $+33.5 \pm 3.8\ \kms$ & $+34.0 \pm 3.9\ \kms$ \\
$V$          & $-20.7 \pm 4.0\ \kms$ & $-20.0 \pm 3.9\ \kms$ & $-20.5 \pm 4.0\ \kms$ \\
$W$          & $-2.5 \pm 1.7\ \kms$  & $-2.4 \pm 1.7\ \kms$ & $-2.5 \pm 1.7\ \kms$ \\
\enddata

\end{deluxetable*}

The refined value of $T_1$ reported here differs by approximately $35\
{\rm K}$ from the assumed value in \S \ref{ebop_lc_sect} used to
determine limb darkening coefficients.  This discrepancy is $<
1\sigma$, and should have a negligible effect on the parameters
determined from the light curve analysis, so we elected not to repeat
this using the updated value of $T_1$ given the computational expense
involved.

Given the full system kinematic information (position, proper motions,
$\gamma$ velocity and parallax from the EB analysis) we can also
calculate the $UVW$ components of the space velocity relative to the
Sun.  We use the method of \citet{johnson1987}, but adopt the definition
that positive $U$ values are away from the Galactic center (a left-handed
coordinate system) for better consistency with the literature.  These are
included in Table \ref{kinparams}.  Comparing to the velocity
ellipsoids derived by \citet{chiba2000}, GJ~3236 appears to be a
member of the Galactic disc, and lies within the bounds of the young
Galactic disc population as defined by \citet{leggett1992}.  Comparing
to the $UVW$ space motions of various solar neighborhood moving groups
(e.g. \citealt{soderblom1993}), the velocities are
consistent with with those of the Hyades moving group, which has
$(U,V,W) = (+38 \pm 6,-17 \pm 6,-11 \pm 12)\ \kms$ \citep{zhao2009}.
Despite this possible membership in the Hyades group, we note that the
dispersion in stellar parameters for the group members does not
provide a very useful constraint on age or metallicity for the present
system.  \citet{zhao2009} obtain a metallicity of ${\rm [Fe/H]} = -0.09
\pm 0.17$ for the Hyades group.

\section{Discussion}
\label{disc_sect}

Figures \ref{mrrel} and \ref{mtrel} show the position of GJ~3236
relative to other known objects and theoretical models of the M-dwarf
mass-radius and mass-effective temperature relationships.

\begin{figure}
\centering
\includegraphics[angle=270,width=3.3in]{f4.eps}
\caption{Mass-radius relation for eclipsing double-lined
  spectroscopic binary systems with one or more components below $0.4\
  \msun$.  Black points with error bars show the two components of the
  GJ~3236 system, for all three solutions reported in Table
  \ref{params}.  The gray points are known systems from the literature
  (\citealt{morales2009}; \citealt{ribas2003}; \citealt{vaccaro2007};
  \citealt{blake2008}).  The black line shows the theoretical
  mass-radius relationship from \citet{bcah98} for an age of $1\ {\rm 
  Gyr}$.}
\label{mrrel}
\end{figure}

\begin{figure}
\centering
\includegraphics[angle=270,width=3.3in]{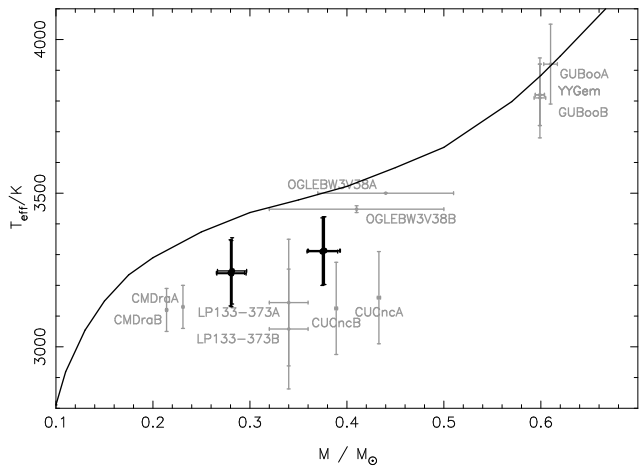}
\caption{As Figure \ref{mrrel}, but showing the mass-effective
  temperature relationship.  Again, all three solutions are shown for
  GJ~3236, but the differences are relatively minor in this plane.  We
  also include several more massive M-dwarf EBs in this figure for
  comparison: YY~Gem \citep{torres2002}, GU~Boo
  \citep{lopezmorales2005}, and OGLE BW3 V38 \citep{maceroni2004}.
  The black line shows the theoretical relationship from
  \citet{bcah98} for an age of $1\ {\rm Gyr}$.} 
\label{mtrel}
\end{figure}

Although many of the best-observed systems are found to exhibit larger
radii than the models predict, the slope of the mass-radius relation
between the components of each system (or equivalently, the ratio of
the component radii) is generally found to agree well between the
observations and the theoretical models.  Examining the positions of
our solutions for GJ~3236, our models 1 and 3 for the out of eclipse
modulations are reasonably consistent with this expectation, whereas
the slope defined by the components for solution 2 is substantially
different.  Although this indicates the assumptions used in model 2
may be unrealistic, the present observational data for GJ~3236 do not
distinguish between the three possibilities.  Using this as a
constraint in the modeling would be dangerous because the results
would then no longer yield a completely independent test of the
theoretical models, so instead we suggest that a better solution to
the issue would be obtaining improved multi-band photometry to
better-constrain the spot parameters.

Therefore, within the present observational uncertainties, the radii
of GJ~3236 appear to be consistent with the theoretical predictions at 
approximately $1 \sigma$, when we account for the systematic
uncertainties represented by the three solutions we have presented.
Furthermore, the central values appear to lie above the theoretical
curve.  This is largely consistent with the results for the well-known
systems in the literature, where the radii are typically found to be
$10-15\%$ larger than the theoretical predictions.

In the effective temperature versus mass plane (Figure \ref{mtrel}),
the effective temperatures of the components of GJ~3236 are found to
be cooler than the models predict, by approximately $2 \sigma$.  This
is in good agreement with many of the well-known systems in the
literature as shown in the figure, such as CM~Dra.

Compared to the well-known systems CM~Dra and CU~Cnc, GJ~3236 has a
shorter orbital period, and the H$\alpha$ emission in the spectroscopy
and out-of-eclipse modulations we observe in the light curve are
indicative of high activity levels, with both the photometric period
and the spectroscopic line broadening apparently consistent with the
stellar spin being synchronized to the binary orbit, as expected from
tidal theory (e.g. \citealt{zahn1977}).

\citet{chabrier2007} propose two hypotheses to explain the observed
radius discrepancies between theoretical models and eclipsing
binaries: (1) that the inflated radii result from reduced convective
efficiency due to high rotation rates and large magnetic fields; or
(2) that magnetic spot coverage of the surfaces leads to reduced heat
flux, and thus larger radii and cooler effective temperatures.
Hypothesis (2) has been found to be consistent with some of the
well-known eclipsing binary systems, e.g. YY~Gem
\citep{vangent1926,joy1926,torres2002}, where the radius discrepancy
(and the discrepancy in effective temperature) can be explained by the
presence of starspots covering approximately $50\%$ of the stellar
surface (if the spots are cooler than the photosphere by $15\%$;
\citealt{ribas2008}).  GJ~3236 shows out of eclipse modulations of
comparable amplitude to YY~Gem, a compatible radius discrepancy within
the present observational errors, and similarly has effective
temperatures somewhat cooler than the models predict, so this is an
attractive hypothesis.  By obtaining precise, multi-band photometry
covering the entire orbital phase, it may be possible to constrain the
spot temperatures, and hence observationally test this argument.

The main limitations in the orbital and geometric modeling of the
system arise from the lack of out-of-eclipse data in multiple
passbands, which leads to large errors ($4-5\%$) on the radii
when we take the degeneracies in the spot configuration into account.
The precision of the mass estimates ($5\%$) is limited by the
error in the radial velocity measurements, and an important
contribution to the uncertainties in the radii will be from the
spectroscopic light ratio if the uncertainty in spot parameters can be
resolved.  Further high-precision photometric measurements and
high-resolution spectroscopy are therefore needed, and combined with
careful analysis these should allow the precision of the mass and
radius measurements for this system to be improved, potentially to
beyond the $2\%$ level as for the well-known systems CM~Dra and
YY~Gem.

\acknowledgments The MEarth team gratefully acknowledges funding from
the David and Lucile Packard Fellowship for Science and Engineering
(awarded to DC).  This material is based upon work supported by the
National Science Foundation under grant number AST-0807690.  SQ and
DWL acknowledge support from the NASA Kepler mission under cooperative
agreement NCC2-1390.  GT acknowledges partial support from the NSF
through grant AST-0708229.  We thank Eric Mamajek for helpful
discussions regarding solar neighborhood kinematics, and the
``astro-comb'' team at CfA: Claire Cramer, David Phillips, Alexander
Glenday, Chih-Hao Li, Dimitar Sasselov and Ronald Walsworth, for their
generous donation of engineering time to obtain spectra of our
target.  The referee is thanked for comments that helped to improve
the manuscript.

This research has made extensive use of data products from the Two
Micron All Sky Survey, which is a joint project of the University of
Massachusetts and the Infrared Processing and Analysis Center /
California Institute of Technology, funded by NASA and the NSF, NASA's
Astrophysics Data System (ADS), and the SIMBAD database, operated at
CDS, Strasbourg, France.  The Digitized Sky Surveys were produced at
the Space Telescope Science Institute under U.S. Government grant NAG
W-2166. The images of these surveys are based on photographic data
obtained using the Oschin Schmidt Telescope on Palomar Mountain and
the UK Schmidt Telescope. The plates were processed into the present
compressed digital form with the permission of these institutions.
The Peters Automated Infrared Imaging Telescope (PAIRITEL) is operated
by the Smithsonian Astrophysical Observatory (SAO) and was made
possible by a grant from the Harvard University Milton Fund, the
camera loan from the University of Virginia, and the continued support
of the SAO and U.C. Berkeley. The PAIRITEL project and JSB are further
supported by NASA/Swift Guest Investigator Grant NNG06GH50G. We
thank M. Skrutskie for his continued support of the PAIRITEL project.

\end{document}